\documentclass[aps,pre,preprint,showpacs,superscriptaddress]{revtex4-1}
\usepackage[utf8]{inputenc}
\usepackage[english]{babel}
\usepackage{amsmath}
\usepackage{amsfonts}
\usepackage{amssymb}
\usepackage[left=2cm,right=2cm,top=2cm,bottom=2cm]{geometry}
\usepackage{graphicx}
\graphicspath{{PDF_figs/}}
\usepackage{color}

\begin{document}


\title{Electro-osmosis in nematic liquid crystals}


\author{O.~M.~Tovkach}
\email[]{otovkach@uakron.edu}
\affiliation{Department of Mathematics, The University of Akron, Akron, OH 44325, USA}
\affiliation{Bogolyubov Institute for Theoretical Physics, NAS of Ukraine, Metrologichna 14-b, Kyiv 03680, Ukraine}
\author{M.~Carme~Calderer}
\email[]{calde014@umn.edu}
\affiliation{School of Mathematics, University of Minnesota, Minneapolis, MN 55455, USA}
\author{Dmitry~Golovaty}
\email[]{dmitry@uakron.edu}
\affiliation{Department of Mathematics, The University of Akron, Akron, OH 44325, USA}
\author{Oleg~Lavrentovich}
\email[]{olavrent@kent.edu}
\affiliation{Liquid Crystal Institute, Kent State University, Kent, OH 44242, USA}
\author{Noel~J.~Walkington}
\email[]{noelw@andrew.cmu.edu}
\affiliation{Department of Mathematical Sciences, Carnegie Mellon University, Pittsburgh, PA 15213, USA}


\date{\today}

\begin{abstract}
We derive a mathematical model of a nematic electrolyte based on variational formulation of nematodynamics. We verify the model by comparing its predictions to the results of the experiments on the substrate-controlled liquid-crystal-enabled electrokinetics. In the experiments a nematic liquid crystal confined to a thin planar cell with surface-patterned anchoring conditions exhibit electro-osmotic flows along the ``guiding rails'' imposed by the spatially varying director. Extending our previous work, we consider a general setup which incorporates dielectric anisotropy of the liquid-crystalline matrix and the full set of nematic viscosities. 
\end{abstract}


\maketitle

\section{Introduction}

With a rapid development of micro- and nanofluidics, a significant effort to understanding electrokinetics has been made in both fundamental and applied science \cite{Ramos, Morgan}. One usually distinguishes between two types of electrokinetic phenomena: an electrically driven transport of particles in a fluid (electrophoresis) and electrically driven flows of fluids with respect to their containers (electro-osmosis). A necessary prerequisite for either of these phenomena to occur is a spatial separation of electric charges. In classical electrokinetics, the separation is achieved through the formation of electric double layers at the solid-fluid interface \cite{Russel}. In the case of electrically neutral but highly polarizable surfaces, the charges can be separated by the applied field. This is the so-called induced-charge electro-osmosis \cite{Bazant}. Once separated, under the action of the field, the charges are forced to move, thus creating a flow of the fluid.

An alternative approach that does not require a solid component is to employ an anisotropic fluid as an electrolyte. The anisotropy makes it possible to separate oppositely charged ions via inhomogeneities of the medium, giving rise to nonlinear electrokinetics. In particular, recent experiments \cite{Hernandez_1, Hernandez_2, Lavrentovich_Nature, Lazo_1, Lazo_2, Sasaki} demonstrate that in nematic liquid crystals, the velocities of the flows are quadratic in the field strength, i.e. do not depend on the field's polarity. This feature enables electrokinetic transport driven by an alternating current and allows one to overcome many technological barriers.

Of particular interest from a theoretical point of view are the experiments on the so-called substrate-controlled liquid-crystal-enabled electrokinetics reported in \cite{Peng_pattern}.
The authors used as an electrolyte a nematic liquid crystal confined to a thin planar cell with surface-patterned anchoring conditions and observed electro-osmotic flows along the ``guiding rails'' imposed by the spatially varying director.
This setup is probably the simplest to analyze as the director patterns in the experiments were periodic, well defined (no topological defects), and homogeneous in the direction of the applied electric field. A mathematical model of this experiment was considered in \cite{Carme} for the specific case of isotropic viscosity and dielectric permittivity of the nematic. In this paper, we propose a more general theory which incorporates the full set of nematic viscosities as well as dielectric anisotropy of the liquid-crystalline matrix that is also expected to trigger electrokinetic flows \cite{Lazo_2,Peng_pattern}.

But before proceeding to this illustrative example in the second part of the paper, we rederive a general system of equations governing electro-osmotic flows in nematic electrolytes. An alternative derivation can be found in \cite{Carme}. Inspired by ideas \cite{Leslie_1, Leslie_2, Walkington}, the authors established a system of governing equations from the local form of balance of linear and angular momentum. Here we arrive at the same results in a more formal, but probably more efficient manner following a variational formulation of nematodynamics suggested in \cite{Sonnet_tensor, Sonnet_dissipative}.

\section{Principle of minimum energy dissipation}

In classical mechanics, irreversible dynamics of a system can be described by means of a Rayleigh dissipation function $\mathcal{R}=\frac{1}{2}\xi_{ij}\dot{q}_i\dot{q}_j$ quadratic in generalized velocities $\dot{q}=(\dot{q}_1, ..., \dot{q}_M)$  (summation over repeated subscripts is implied hereafter).
The basic idea is to balance frictional and conservative forces in Lagrange's dynamical equations 
\begin{equation}\label{Lagrange_eq}
\frac{d}{dt}\frac{\partial \mathcal{L}}{\partial \dot{q}_m} -\frac{\partial \mathcal{L}}{\partial q_m} +\frac{\partial \mathcal{R}}{\partial \dot{q}_m} = 0,
\end{equation}
where $q=(q_1, ..., q_M)$ are generalized coordinates conjugated with the velocities $\dot{q}$ and $\mathcal{L}=\frac{1}{2}a_{ij}(q)\dot{q}_i\dot{q}_j-\mathcal{U}(q)$ is the Lagrangian of the system, defined as the difference between the kinetic energy $\frac{1}{2}a_{ij}(q)\dot{q}_i\dot{q}_j$ and the potential energy $\mathcal{U}(q)$. In what follows, we assume that the matrices $\left(\xi_{ij}\right)$ and $\left(a_{ij}\right)$ are symmetric.

Similarly to their non-dissipative counterparts,  Eqs.~\eqref{Lagrange_eq} can be recast into a variational problem as their solutions provide a critical point of the functional 
$$\int_{\Omega} d^3r\left\{ \dot{\mathcal{E}}+\mathcal{R} \right\}$$
with respect to a special class of variations $\delta\dot{q}$ of the generalized velocities $\dot{q}$. 
Here $\Omega\subset\mathbb R^3$ is the region occupied by the system, $\mathcal{E}=\mathcal{L}+2\mathcal{U}$ is the total energy and the superimposed dot (as well as $\frac{d}{dt}$) denotes the total or material time derivative.
Unlike Hamilton's principle of stationary action, the current approach ``freezes'' both the configuration $q$ and the generalized forces $X_m:=\frac{d}{dt}\frac{\partial\mathcal{L}}{\partial \dot{q}_m}-\frac{\partial\mathcal{L}}{\partial q_m}$, $m=1,\ldots,M$ acting on the system at a given moment of time. The state of the system is then varied by imposing arbitrary instantaneous variations $\delta\dot{q}$ of the velocities $\dot{q}$. Note that here $\delta\dot{q}$ {\bf are not} time derivatives of $\delta q$. Similarly, variations $\delta\ddot{q}$ should not be confused with time derivatives of the $\delta\dot{q}$, but have to be chosen instead so as to keep the generalized forces $X_m,\ m=1,\ldots,M$ unaltered \cite{Sonnet_book}.
Then, by using the product rule and relabeling, we indeed have
\begin{multline}
\frac{\delta}{\delta \dot{q}_m} \int_{\Omega} d^3r \left\{ \dot{\mathcal{E}} +\mathcal{R} \right\} 
\frac{\delta}{\delta \dot{q}_m} \int_{\Omega} d^3r \left\{ a_{ij}\ddot{q}_j\dot{q}_i +\frac{1}{2}\frac{\partial a_{ij}}{\partial q_k}\dot{q}_k\dot{q}_j\dot{q}_i +\frac{\partial \mathcal{U}}{\partial q_i}\dot{q}_i +\mathcal{R} \right\} \\
= \frac{\delta}{\delta \dot{q}_m} \int_{\Omega} d^3r \left\{ \left[\frac{d}{dt}\left( a_{ij}\dot{q}_j \right) -\frac{1}{2}\frac{\partial a_{kj}}{\partial q_i}\dot{q}_k\dot{q}_j +\frac{\partial \mathcal{U}}{\partial q_i}\right]\dot{q}_i +\mathcal{R} \right\} =\frac{\delta}{\delta \dot{q}_m} \int_{\Omega} d^3r \left\{ X_i\dot{q}_i +\mathcal{R} \right\} \\
=X_m+\frac{\partial \mathcal{R}}{\partial \dot{q}_m}=\frac{d}{dt} \frac{\partial \mathcal{L}}{\partial \dot{q}_m} -\frac{\partial \mathcal{L}}{\partial q_m} +\frac{\partial \mathcal{R}}{\partial \dot{q}_m},
\end{multline}
for every $m=1,\ldots,M$.
Hence, the Euler-Lagrange equations
\begin{equation}\label{Principle}
\frac{\delta}{\delta \dot{q}} \int_{\Omega} d^3r \left\{ \dot{\mathcal{E}} +\mathcal{R}\right\} = 0
\end{equation}
are identical to the generalized equations of motion \eqref{Lagrange_eq} and thus govern dynamics of a dissipative mechanical system. Since the conservative forces  are assumed to be fixed here and $\mathcal{R}$ is a positive-definite function, the equations \eqref{Principle} yield a minimum of energy dissipation \cite{Sonnet_tensor, Sonnet_dissipative}.
It is worth noting that for overdamped systems---where $\ddot{q}=0$---this principle of minimum energy dissipation is equivalent to the Onsager's variational approach \cite{Doi}.  

\section{Nematic electrolyte}

In this section, we apply the principle \eqref{Principle} to a nematic electrolyte subject to an external electric field.
It was shown earlier that under an appropriate choice of the generalized velocities this framework is capable of reproducing the classical Ericksen-Leslie equations of nematodynamics \cite{Sonnet_tensor, Sonnet_dissipative}.
Below we demonstrate that it can be extended so as to take into account the presence of an ionic subsystem.

\subsection{Energy of the system}

Consider a nematic liquid crystal that contains an ideal gas of $N$ species of ions with valences $z^{\alpha}$ at concentrations $c^{\alpha}$, where $1\leq\alpha\leq N$.
Assuming that the ions do not interact with the liquid crystal one can write the density of the ionic subsystem energy in the form of entropic and Coulombic contributions  
\begin{equation}\label{E_ion}
\mathcal{E}_{ion} = k_{B} \Theta \sum_{\alpha=1}^{N} c^{\alpha} \ln c^{\alpha} +\sum_{\alpha=1}^{N} e c^{\alpha} z^{\alpha} \Phi,
\end{equation}
where $k_B$ and $\Theta$ stand for the Boltzmann constant and the absolute temperature, respectively, $\Phi$ denotes the electric potential.
Under the action of the field, the ions move with velocities $\mathbf{u}^{\alpha}$ which satisfy the continuity equations 
\begin{equation}\label{Charge_conservation}
\frac{\partial c^{\alpha}}{\partial t} + \nabla\cdot (c^{\alpha}\mathbf{u}^{\alpha}) = 0.
\end{equation}

Nematics themselves are anisotropic ordered fluids.
A typical member of their family consists of elongated molecules whose local orientation can be described by a coarse-grained vector field $\mathbf{n}\equiv -\mathbf{n}$ with non-polar symmetry, the director.
This unit-length field allows us the represent the elastic energy of the liquid crystal in the Oseen-Frank form 
\begin{equation}\label{Frank_energy}
\mathcal{E}_{OF} = \frac{K_1}{2}\left(\nabla\cdot\mathbf{n}\right)^2 +\frac{K_2}{2}\left( \mathbf{n}\cdot\left[\nabla\times\mathbf{n}\right] \right)^2 +\frac{K_3}{2}\left(\mathbf{n}\times\left[\nabla\times\mathbf{n}\right]  \right)^2,
\end{equation} 
where $K_1$, $K_2$ and $K_3$ are positive, non-zero constants and pure divergence terms are omitted.
 
In order to take into account the coupling between the electric field $\mathbf{E}=-\nabla\Phi$ and the director, we have to supplement the potential energy of the nematic by 
\begin{equation}\label{E_E}
\mathcal{E}_{E} = -\frac{1}{2}\mathbf{D}\cdot\mathbf{E},
\end{equation}
where $\mathbf{D}$ denotes the electric displacement.
It should be noted that care must be taken in dealing with the electric field in this problem.
The field is substantially nonlocal, that is, its changes can affect the system even if they occur outside the region $\Omega$ occupied by the system.
In order to avoid dealing with the field outside of $\Omega$, we assume that the system under investigation is surrounded by conductors that are held at a prescribed potential $\Phi_{\partial\Omega}$.
Then the electric field exists in $\Omega$ only so that
\begin{equation}
D_i = \varepsilon_0\varepsilon_{ij}E_j  = \varepsilon_0(\varepsilon_{\perp}\delta_{ij}+\Delta\varepsilon n_i n_j)E_j,
\end{equation}
where $\Delta\varepsilon=\varepsilon_{\|}-\varepsilon_{\perp}$, $\varepsilon_{\perp}$ and $\varepsilon_{\|}$ are dielectric permittivities perpendicular and along the director, respectively, measured in units of the vacuum permittivity $\varepsilon_0$.
Following Maxwell, the electric displacement $\mathbf{D}$ obeys 
\begin{equation}\label{Maxwell_eq}
\nabla\cdot\mathbf{D}=\sum_{\alpha=1}^{N} e c^{\alpha} z^{\alpha}.
\end{equation}

Thus, neglecting inertia of the director rotation $(\ddot{\mathbf{n}}=0)$, one can write the total energy per unit volume of the system in the form
\begin{equation}\label{Total_energy}
\mathcal{E}= \frac{1}{2}\rho v_i v_i +\mathcal{E}_{OF} +\mathcal{E}_{E} +\mathcal{E}_{ion}
\end{equation}
with $\rho=\text{const}$ being the nematic mass density and $\mathbf{v}$ the velocity of its flow which we assume to be incompressible, $\nabla\cdot\mathbf{v}=0$.

\subsection{Dissipation function}

Within the current framework, the dissipation function has to be frame-indifferent, positive-definite and quadratic in the generalized velocities. 
As we mentioned above, the principle of minimum energy dissipation results in the correct nematodynamics when those velocities are $\mathbf{v}$ and $\dot{\mathbf{n}}$. 
Then the dissipation function of a nematic liquid crystal can be written in the following form \cite{Sonnet_dissipative}
\begin{equation}\label{R_nem}
2\mathcal{R}_{nem} = \gamma_1 \mathring{n_i}^2 +2\gamma_2 \mathring{n_i} \mathsf{A}_{ij}n_j +\gamma_3(\mathsf{A}_{ij}n_j)^2 +\gamma_4(\mathsf{A}_{ij})^2+\gamma_5(n_i \mathsf{A}_{ij} n_j)^2,
\end{equation}
where $\mathsf{A}_{ij}=\frac{1}{2}(\partial_j v_i +\partial_i v_j)$ represent the symmetric part of the velocity gradient and $\mathring{n_i}=\dot{n_i}-\frac{1}{2}(\partial_j v_i -\partial_i v_j)n_j$. 
The Lie derivative of the director, $\mathring{\mathbf{n}}$, gives its rate of change relative to a flow vorticity. 
Below we will see that $\mathcal{R}_{nem}$ indeed yields the well-known nematic viscous stress, provided that the $\gamma$s in \eqref{R_nem} are related to Leslie's viscosities $\alpha$s via the following
\begin{equation}\label{Gammas}
\begin{split}
\alpha_1=\gamma_5 \quad \alpha_2=\frac{1}{2}(\gamma_2 -\gamma_1) \quad \alpha_3=\frac{1}{2}(\gamma_2 +\gamma_1)\\
\alpha_4=\gamma_4 \quad \alpha_5=\frac{1}{2}(\gamma_3 -\gamma_2) \quad \alpha_6=\frac{1}{2}(\gamma_3 +\gamma_2).
\end{split}
\end{equation}
Note that under these circumstances, the Parodi's relation, $\alpha_6-\alpha_5=\alpha_2+\alpha_3$, is automatically satisfied.
Besides, positive definiteness of $\mathcal{R}_{nem}$ requires that \cite{Parodi}
\begin{equation}\label{Visc_constr}
\begin{split}
\alpha_4>0, \quad \alpha_3>\alpha_2, \quad 2\alpha_4+\alpha_5+\alpha_6>0,\\
\alpha_1+\alpha_4+\alpha_5+\alpha_6>0, \quad (\alpha_3-\alpha_2)(2\alpha_4+\alpha_5+\alpha_6)>(\alpha_6-\alpha_5)^2.
\end{split}
\end{equation}

For the system under consideration, additional degrees of freedom  are brought in by the ions.
Although they do not interact with the nematic via potential forces, their motion with respect to the liquid crystal contributes to the dissipation \cite{Carme}
\begin{equation}\label{R_ion}
2\mathcal{R}_{ion} = k_B\Theta \sum_{\alpha_1}^{N} c^{\alpha}(\mathsf{D}_{ij}^{\alpha})^{-1}(u_i^{\alpha}-v_i)(u_j^{\alpha}-v_j).
\end{equation}
Here the diffusion matrix $\mathsf{D}_{ij}^{\alpha}$ reflects the anisotropy of the liquid crystal conductivity.
Generally, mobilities of ions along and perpendicular to the director $\mathbf{n}$ are different. 
Apparently, \eqref{R_ion} is indeed the dissipation function if $\mathbf{u}^{\alpha}$ with $1\leq\alpha\leq N$ are also treated as the generalized velocities.

Thus, the total energy dissipation in the system is the sum $\mathcal{R}=\mathcal{R}_{nem}+\mathcal{R}_{ion}$.

\subsection{Governing equations}

Once the energy $\mathcal{E}$, the dissipation $\mathcal{R}$, and the generalized velocities of the system are specified, we are in a position to derive equations describing electro-osmotic flows in nematics.
The equations are implicitly given by
\begin{equation}\label{EL_eq}
\begin{split}
\frac{\delta}{\delta \mathbf{v}}\int_{\Omega} d^3r \left\{ \dot{\mathcal{E}}+\mathcal{R} -p (\partial_i v_i) -\Lambda n_i\dot{n_i}  \right\}=0, \\
\frac{\delta}{\delta \dot{\mathbf{n}}}\int_{\Omega} d^3r \left\{ \dot{\mathcal{E}}+\mathcal{R} -p (\partial_i v_i) -\Lambda n_i\dot{n_i}  \right\}=0, \\
\frac{\delta}{\delta \mathbf{u}^{\alpha}}\int_{\Omega} d^3r \left\{ \dot{\mathcal{E}}+\mathcal{R} -p (\partial_i v_i) -\Lambda n_i\dot{n_i}  \right\}=0,
\end{split}
\end{equation}
where two Lagrange multipliers, $p$ and $\Lambda$, associated, respectively, with the flow incompressibility and the director's unit length appear.

But before proceeding to  an explicit form of \eqref{EL_eq}, let us address the boundary conditions for our problem.
Rigorously speaking, the principle of minimum energy dissipation makes it possible to derive appropriate boundary conditions directly from \eqref{Principle} (natural boundary conditions).
Here we, however, impose Dirichlet conditions on the system's boundary $\partial\Omega$.
In particular,
\begin{equation}
\mathbf{v}=0,\quad \dot{\mathbf{n}}=0,\quad \mathbf{u}^{\alpha}=0 \quad \text{on } \partial\Omega.
\end{equation}
Such a choice slightly simplifies further consideration and should correspond to a majority of experimental setups.

Given these preliminary arguments, consider again Eq.~\eqref{EL_eq}.
First, calculate the rate of energy change. 
For the sake of clarity, we divide this process into the following steps
\begin{multline}\label{E_nem_dot}
\frac{d}{d t} \int_{\Omega} d^3r \left\{ \frac{1}{2} \rho \mathbf{v}^2 +\mathcal{E}_{OF}\left(\mathbf{n}, \nabla\mathbf{n}\right) \right\} =  \\
= \int_{\Omega} d^3r \left\{\left[ \rho\dot{v_k} +\partial_j\left(\frac{\partial \mathcal{E}_{OF}}{\partial (\partial_j n_i)}(\partial_k n_i)\right) \right] v_k +\left[\frac{\partial \mathcal{E}_{OF}}{\partial n_i} -\partial_j\left(\frac{\partial \mathcal{E}_{OF}}{\partial (\partial_j n_i)}\right) \right]\dot{n_i}   \right\} 
\end{multline}
Similarly, by means of the identity $\dot{\left(\partial_i\Phi\right)} = \partial_i\dot{\Phi} -\left(\partial_i v_k\right)\left(\partial_k\Phi\right)$ we have
\begin{equation}
\frac{d}{dt} \int_{\Omega} d^3r \mathcal{E}_{E}(\mathbf{n}, \nabla\Phi) = 
\int_{\Omega} d^3r \left\{ \frac{\partial \mathcal{E}_{E}}{\partial n_i}\dot{n_i} +\frac{\partial \mathcal{E}_{E}}{\partial (\partial_i \Phi)}(\partial_i \dot{\Phi}) -\frac{\partial \mathcal{E}_{E}}{\partial (\partial_i \Phi)}(\partial_k \Phi)(\partial_i v_k)\right\}.
\end{equation}
Recall that $\mathcal{E}_{E} = -\varepsilon_0 (\varepsilon_{\perp}\delta_{ij}+\Delta\varepsilon n_i n_j)(\partial_i \Phi) (\partial_j \Phi)/2$ so that
\begin{equation}
\begin{split}
&\frac{\partial \mathcal{E}_E}{\partial n_i} = -\varepsilon_0 \Delta\varepsilon n_j (\partial_i \Phi) (\partial_j \Phi),\\
&\frac{\partial \mathcal{E}_E}{\partial (\partial_i \Phi)} = -\varepsilon_0 \varepsilon_{ij} (\partial_j \Phi).
\end{split}
\end{equation}
Then
\begin{multline}\label{E_field_dot}
\frac{d}{dt} \int_{\Omega} d^3r \mathcal{E}_{E}(\mathbf{n}, \nabla\Phi) = \\
= \int_{\Omega} d^3r \left\{ (-\varepsilon_0\Delta\varepsilon n_j E_i E_j) \dot{n_i} -(\partial_i D_i) \dot{\Phi} -\partial_i (\varepsilon_0\varepsilon_{ij} E_j E_k) v_k  \right\} +\int_{\partial\Omega} d^2r \left\{ (\nu_i \varepsilon_0\varepsilon_{ij}E_j) \dot{\Phi} \right\}
\end{multline}
Implying that on a conductor-dielectric interface the normal component of the displacement, $D_i\nu_i$, is given by the surface charge density $\sigma$, one sees that the surface integral in \eqref{E_field_dot}
\begin{equation}
\int_{\partial\Omega} d^2r \left\{ (\nu_i \varepsilon_0\varepsilon_{ij}E_j) \dot{\Phi} \right\} = \int_{\partial\Omega} d^2r \nu_i D_i \frac{\partial\Phi_{\partial\Omega}}{\partial t} =\int_{\partial\Omega} d^2r \sigma \frac{\partial \Phi_{\partial\Omega}}{\partial t},
\end{equation}
gives the power of charges located at $\partial\Omega$.
This term can be omitted when $\Phi_{\partial\Omega}$ varies slowly compared to the dynamics given by $\mathbf{v}$, $\mathbf{u}^{\alpha}$ and $\dot{\mathbf{n}}$.

For the ionic subsystem we have
\begin{equation}\label{E_ion_dot}
\frac{d}{dt}\int_{\Omega} d^3r \mathcal{E}_{ion}(c^{\alpha}, \Phi)=\\
\int_{\Omega} d^3r \sum_{\alpha=1}^{N} \left\{(\partial_i \mu^{\alpha}) c^{\alpha} (u_i^{\alpha} -v_i) +e c^{\alpha}z^{\alpha}\dot{\Phi} -\mu^{\alpha}c^{\alpha}(\partial_i v_i) \right\},
\end{equation}
where $\mu^{\alpha} = \frac{\partial \mathcal{E}_{ion}}{\partial c^{\alpha}} = k_{B}\Theta (\ln c^{\alpha} +1) +e z^{\alpha} \Phi$ is identified as the chemical potential of the $\alpha$-th ion species \cite{Eisenberg} and the continuity equation \eqref{Charge_conservation} is used.

Note that $\dot{\mathcal{E}}_{ion}$ includes the term $\sum_{\alpha}ec^{\alpha}z^{\alpha}\dot{\Phi}$ whereas $\dot{\mathcal{E}}_{E}$ contains $-(\partial_i D_i)\dot{\Phi}$.
Obviously, both these terms annihilate and are not present in the total power $\dot{\mathcal{E}}$.
This point deserves a special comment.
Developing the current approach, we initially postulated that the electric field obeys Maxwell's equations.
But now we see that this assumption is not indispensable.
The same equation for $\mathbf{D}$ follows from \eqref{Principle}, provided that $\dot{\Phi}$ is treated as a generalized velocity.
Then
\begin{equation}
\frac{\delta}{\delta \dot{\Phi}}\int_{\Omega} d^3r \left\{ \dot{\mathcal{E}}+\mathcal{R} -p (\partial_i v_i) -\Lambda n_i\dot{n_i}  \right\}=-\partial_i D_i +\sum_{\alpha=1}^{N} ec^{\alpha}z^{\alpha} = 0. 
\end{equation}
Since the present framework deals with the energy of the entire system this derivation properly addresses nonlocality of the field.

Now we can write down the variational derivatives of the dissipation function.
Particularly,
\begin{eqnarray}\label{dR}
\frac{\delta}{\delta \dot{n_i}} \int_{\Omega}d^3r\mathcal{R}= \frac{\partial \mathcal{R}_{nem}}{\partial \mathring{n_i}} = \gamma_1\mathring{n_i} +\gamma_2 \mathsf{A}_{ij}n_j,\label{dRdn}\\
\frac{\delta}{\delta u_i^{\alpha}} \int_{\Omega}d^3r\mathcal{R} = k_B\Theta c^{\alpha}(\mathsf{D}_{ij}^{\alpha})^{-1}(u_j^{\alpha}-v_j),\label{dRdu}\\
\frac{\delta}{\delta v_i} \int_{\Omega}d^3r\mathcal{R}= \frac{\delta}{\delta v_i}\int_{\Omega}d^3r\mathcal{R}_{nem} - k_B\Theta \sum_{\alpha=1}^{N}c^{\alpha}(\mathsf{D}_{ij}^{\alpha})^{-1}(u_j^{\alpha}-v_j).\label{dRdv}
\end{eqnarray}
Keeping in mind the explicit form \eqref{R_nem} of $\mathcal{R}_{nem}$ and relations \eqref{Gammas} for $\gamma$s, one sees that $\frac{\delta}{\delta v_i}\int_{\Omega}d^3r\mathcal{R}_{nem}$ indeed yields divergence of the well-known viscous stress tensor $\mathsf{T}_{ij}^V$ \cite{Leslie_1}
\begin{multline}\label{Viscous_stress}
\frac{\delta}{\delta v_i}\int_{\Omega}d^3r\mathcal{R}_{nem} = -\partial_j \frac{\partial \mathcal{R}_{nem}}{\partial(\partial_j v_i)} =-\partial_j \mathsf{T}_{ij}^V =\\
=-\partial_j \left( \alpha_1 n_i n_j n_k n_l \mathsf{A}_{kl} +\alpha_2 \mathring{n_i}n_j +\alpha_3 n_i\mathring{n_j} +\alpha_4 \mathsf{A}_{ij} +\alpha_5 \mathsf{A}_{ik}n_k n_j +\alpha_6 \mathsf{A}_{kj}n_k n_i \right).
\end{multline}

Thus, it follows from \eqref{E_ion_dot} and \eqref{dRdu} that
\begin{equation}\label{Nernst0_eq}
\frac{\delta}{\delta u_i^{\alpha}}\int_{\Omega} d^3r \left\{ \dot{\mathcal{E}}+\mathcal{R} -p (\partial_i v_i) -\Lambda n_i\dot{n_i}  \right\}= c^{\alpha}\left(\partial_i \mu^{\alpha} +k_B\Theta (\mathsf{D}_{ij}^{\alpha})^{-1}(u_j^{\alpha}-v_j)\right) = 0.
\end{equation}
Combining this with the continuity equation \eqref{Charge_conservation}, we arrive at
\begin{equation}\label{Nernst_eq}
\frac{\partial c^{\alpha}}{\partial t} +\partial_j \left[ c^{\alpha}v_j -\frac{c^{\alpha}}{k_B\Theta}\mathsf{D}_{ij}^{\alpha}(\partial_i \mu^{\alpha}) \right] = 0.
\end{equation}
In the same way, equations \eqref{E_nem_dot}, \eqref{E_field_dot} and \eqref{dRdn} yield
\begin{multline}\label{Leslie_eq}
\frac{\delta}{\delta \dot{n_i}}\int_{\Omega} d^3r \left\{ \dot{\mathcal{E}}+\mathcal{R} -p (\partial_i v_i) -\Lambda n_i\dot{n_i}  \right\}= \\
= \frac{\partial \mathcal{E}_{OF}}{\partial n_i} -\partial_j \left[ \frac{\partial \mathcal{E}_{OF}}{\partial(\partial_j n_i)} \right] -\Lambda n_i +\gamma_1\mathring{n_i} +\gamma_2 \mathsf{A}_{ij}n_j -\varepsilon_0\Delta\varepsilon n_j E_j E_i =0.
\end{multline}
Finally, combining \eqref{E_nem_dot}, \eqref{E_field_dot}, \eqref{E_ion_dot}, \eqref{Viscous_stress} and \eqref{Nernst0_eq} we arrive at
\begin{multline}\label{Navier0_eq}
\frac{\delta}{\delta v_i}\int_{\Omega} d^3r \left\{ \dot{\mathcal{E}}+\mathcal{R} -p (\partial_i v_i) -\Lambda n_i\dot{n_i}  \right\}= \\
= \rho \dot{v_i} +\partial_j \left[ \frac{\partial \mathcal{E}_{OF}}{\partial(\partial_j n_k)}(\partial_i n_k) +p\delta_{ij} -\mathsf{T}_{ij}^V -\varepsilon_0\varepsilon_{jk}E_k E_i \right] =0.
\end{multline}
Recalling that $\nabla\times\mathbf{E}=0$ and $\partial_i \mu^{\alpha} = \partial_i\left[ k_B\Theta (\ln c^{\alpha}+1) \right] +e z^{\alpha}(\partial_i\Phi)$, one can show that
\begin{multline}\label{E_to_p}
-\partial_j\left[\varepsilon_0\varepsilon_{jk}E_k E_i\right] = \\
=\sum_{\alpha=1}^{N}c^{\alpha}(\partial_i \mu^{\alpha}) +\varepsilon_0\Delta\varepsilon n_k E_k (\partial_i n_j) E_j -\partial_i \left[ \varepsilon_0\varepsilon_{\perp} E_k^2 +\varepsilon_0\Delta\varepsilon n_j E_j n_k E_k  +k_B\Theta \sum_{\alpha=1}^{N}c^{\alpha} \right].
\end{multline}
The sum of the gradient term from \eqref{E_to_p}, the Lagrange multiplier $p$ and $\partial_i\left(\mu^{\alpha}c^{\alpha}\right)$ from \eqref{Navier0_eq} can be defined as the total pressure, yielding thus an alternative form 
\begin{equation}\label{Navier_eq}
\rho \dot{v_i} +\partial_j \left[ \frac{\partial \mathcal{E}_{OF}}{\partial(\partial_j n_k)}(\partial_i n_k) +p\delta_{ij} -\mathsf{T}_{ij}^V \right] +\varepsilon_0\Delta\varepsilon n_k E_k (\partial_i n_j) E_j +\sum_{\alpha=1}^{N}c^{\alpha}(\partial_i \mu^{\alpha}) =0.
\end{equation}
of \eqref{Navier0_eq}.
Equations \eqref{Maxwell_eq}, \eqref{Nernst_eq}, \eqref{Leslie_eq} and \eqref{Navier_eq} along with 
the definition of the chemical potential
\begin{equation}
\mu^{\alpha} = \frac{\partial \mathcal{E}_{ion}}{\partial c^{\alpha}} = k_{B}\Theta (\ln c^{\alpha} +1) +e z^{\alpha} \Phi
\end{equation}
and constraints $\nabla\cdot\mathbf{v}=0$,  $\mathbf{n}^2=1$ constitute the full set of equations governing electro-osmosis in nematic liquid crystals,

\begin{equation}\label{The_System}
\begin{cases}
\frac{\partial c^{\alpha}}{\partial t} +\partial_j \left[ c^{\alpha}v_j -\frac{c^{\alpha}}{k_B\Theta}\mathsf{D}_{ij}^{\alpha}(\partial_i \mu^{\alpha}) \right] = 0,\\
\frac{\partial \mathcal{E}_{OF}}{\partial n_i} -\partial_j \left[ \frac{\partial \mathcal{E}_{OF}}{\partial(\partial_j n_i)} \right] -\Lambda n_i +\gamma_1\mathring{n_i} +\gamma_2 \mathsf{A}_{ij}n_j -\varepsilon_0\Delta\varepsilon n_j E_j E_i =0,\\
\rho \dot{v_i} +\partial_j \left[ \frac{\partial \mathcal{E}_{OF}}{\partial(\partial_j n_k)}(\partial_i n_k) +p\delta_{ij} -\mathsf{T}_{ij}^V \right] +\varepsilon_0\Delta\varepsilon n_k E_k (\partial_i n_j) E_j +\sum_{\alpha=1}^{N}c^{\alpha}(\partial_i \mu^{\alpha}) =0,\\
\partial_i \left[ \varepsilon_{\perp}E_j\delta_{ij}  +\Delta\varepsilon n_i n_j E_j\right] = \frac{e}{\varepsilon_0}\sum_{\alpha=1}^{N}c^{\alpha}z^{\alpha},\\
\mu^{\alpha} = \frac{\partial \mathcal{E}_{ion}}{\partial c^{\alpha}} = k_{B}\Theta (\ln c^{\alpha} +1) +e z^{\alpha} \Phi,\\
\partial_i v_i =0,\\
n_i n_i =1.
\end{cases}
\end{equation}

\section{Electro-osmotic flow in a patterned cell}

\begin{figure}
\begin{center}
\includegraphics[width=.3\textwidth]{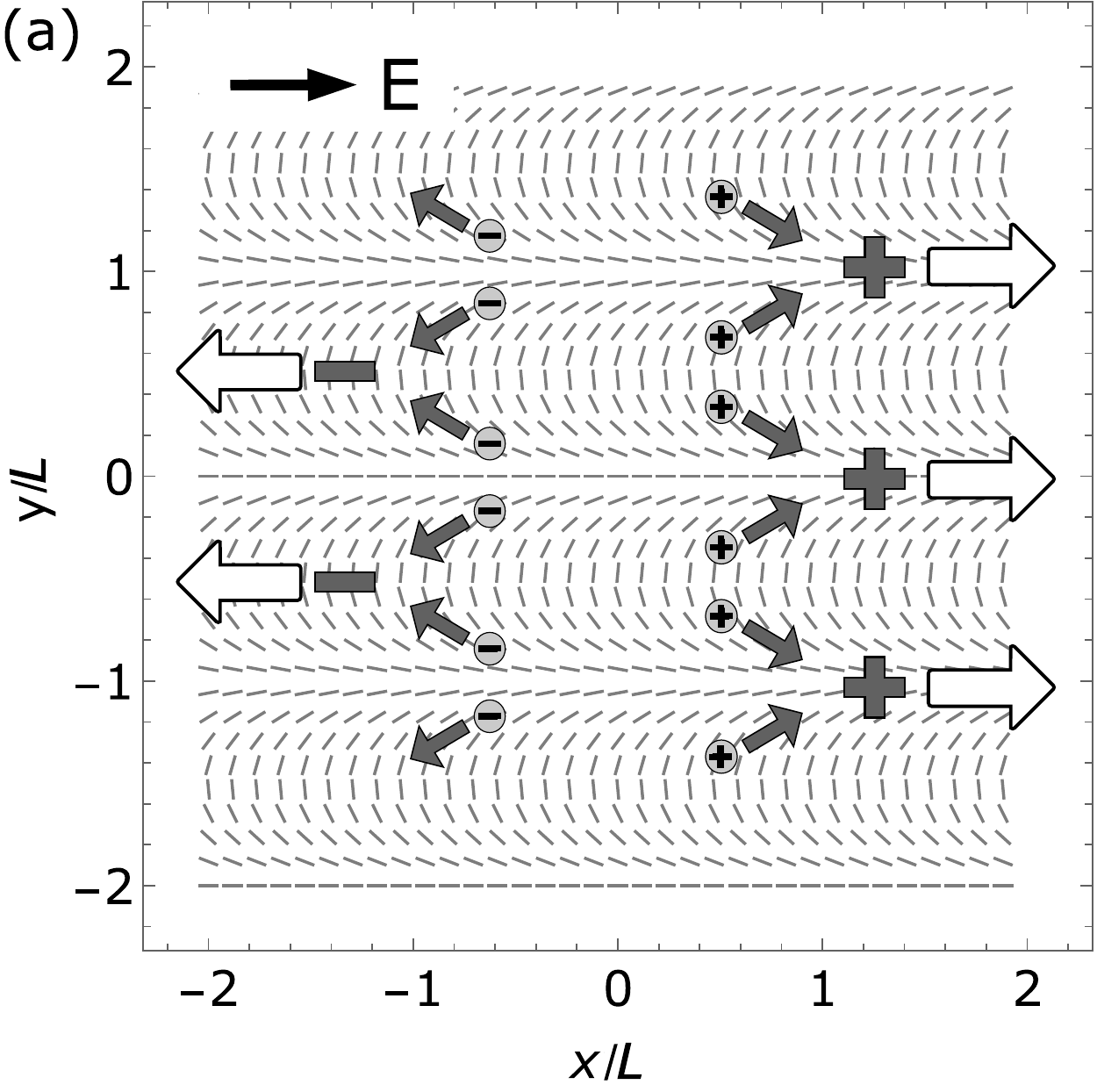}
\includegraphics[width=.3\textwidth]{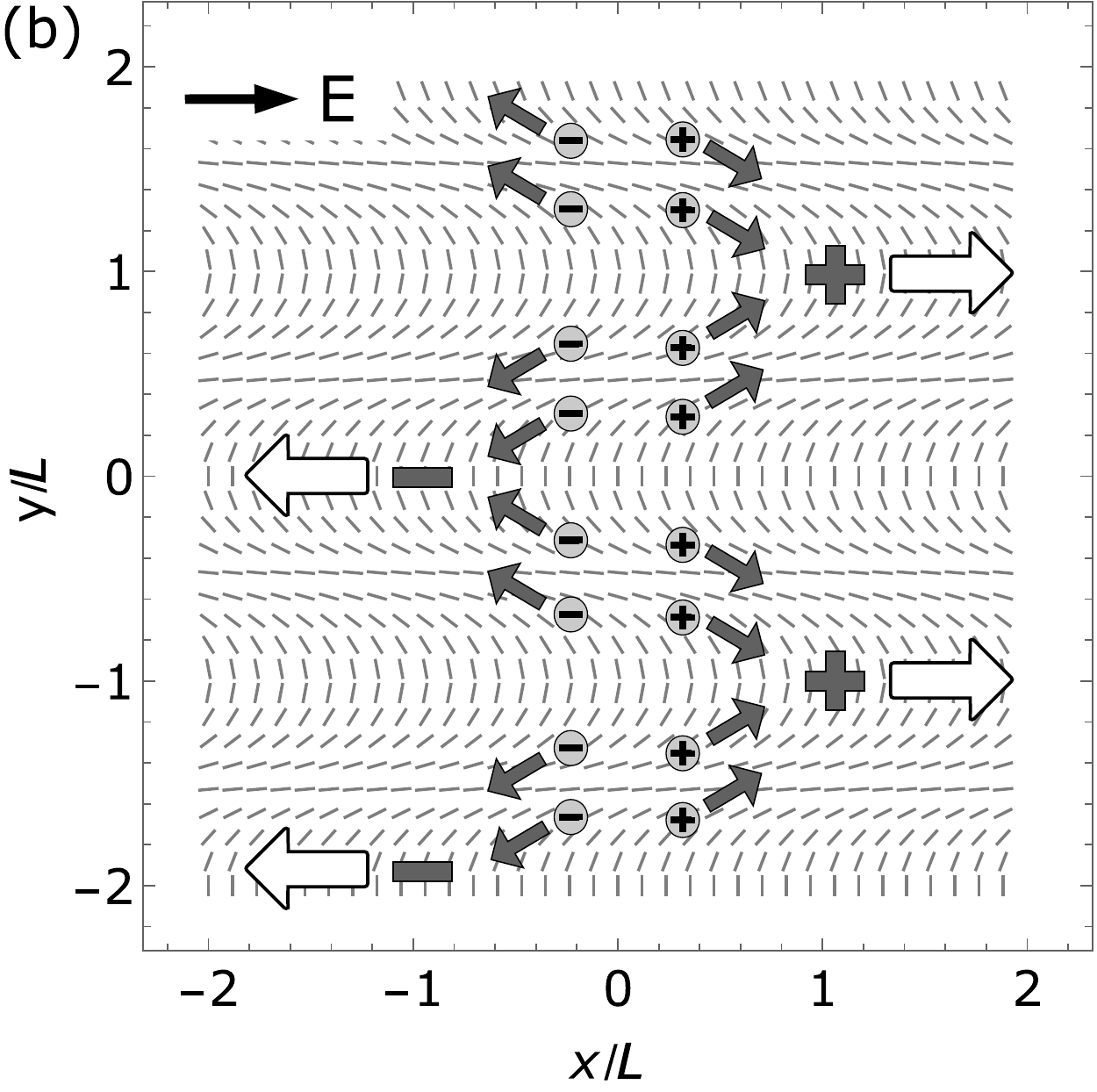}
\includegraphics[width=.3\textwidth]{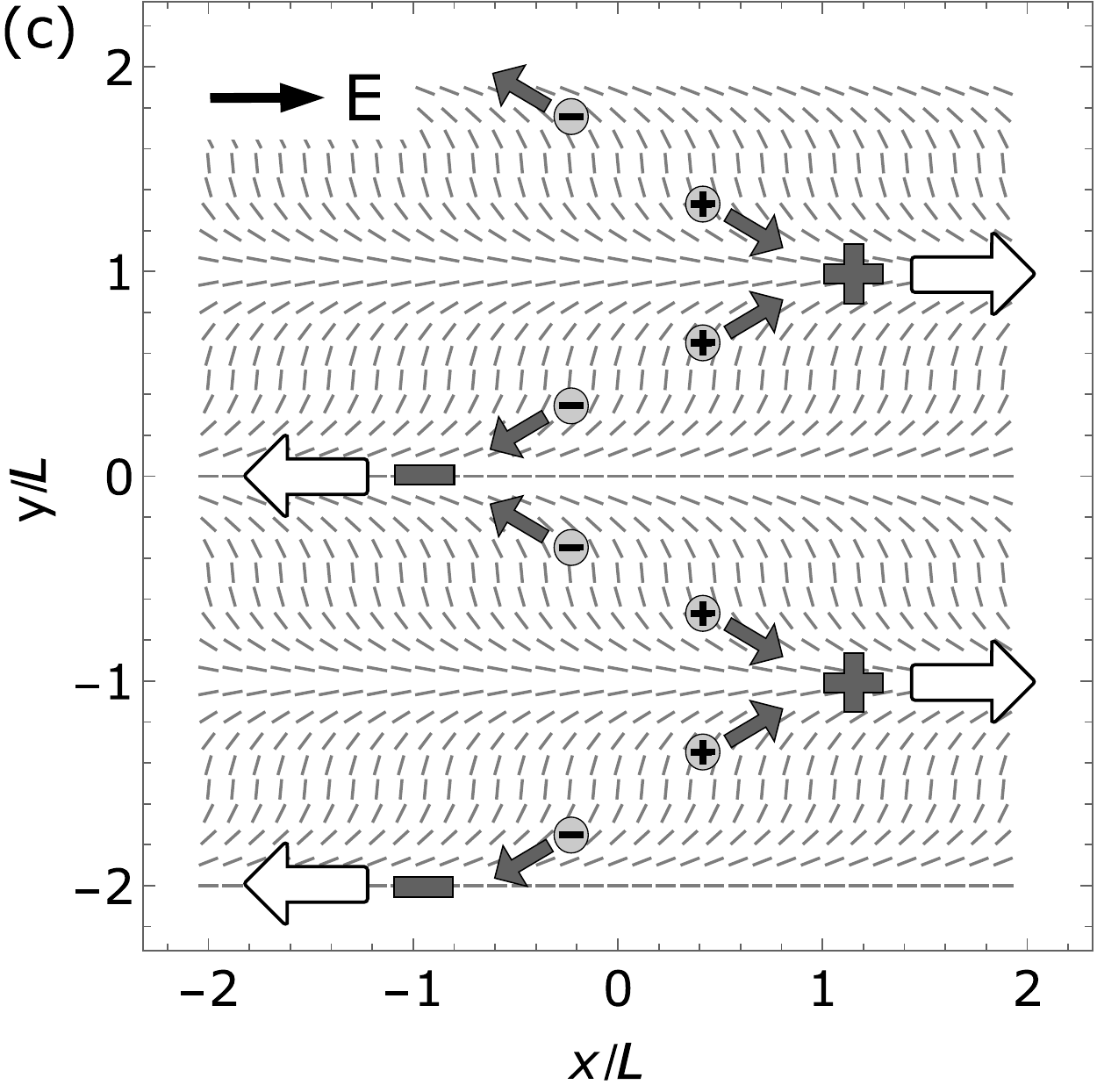}\\
\includegraphics[width=.3\textwidth]{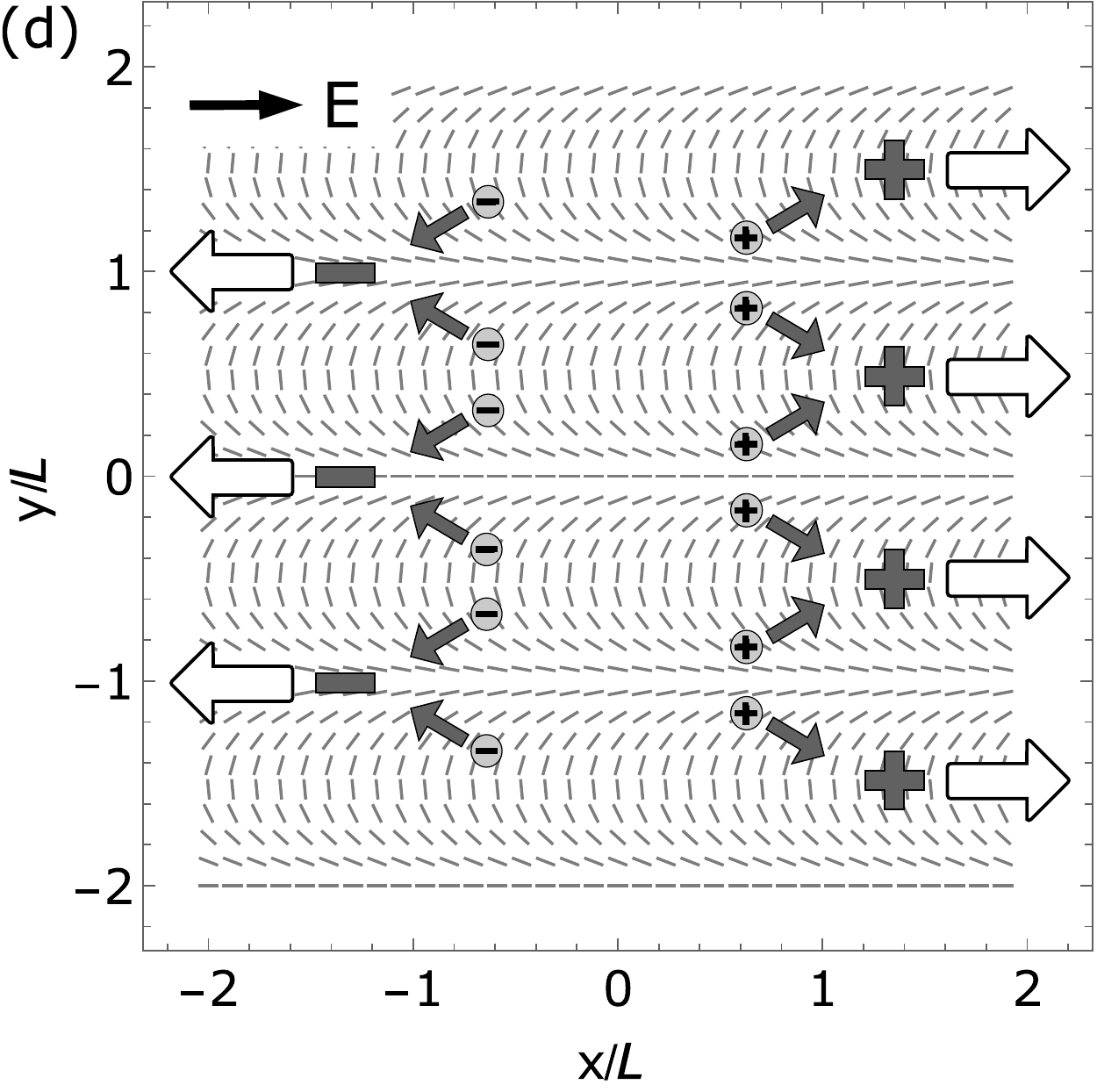}
\includegraphics[width=.3\textwidth]{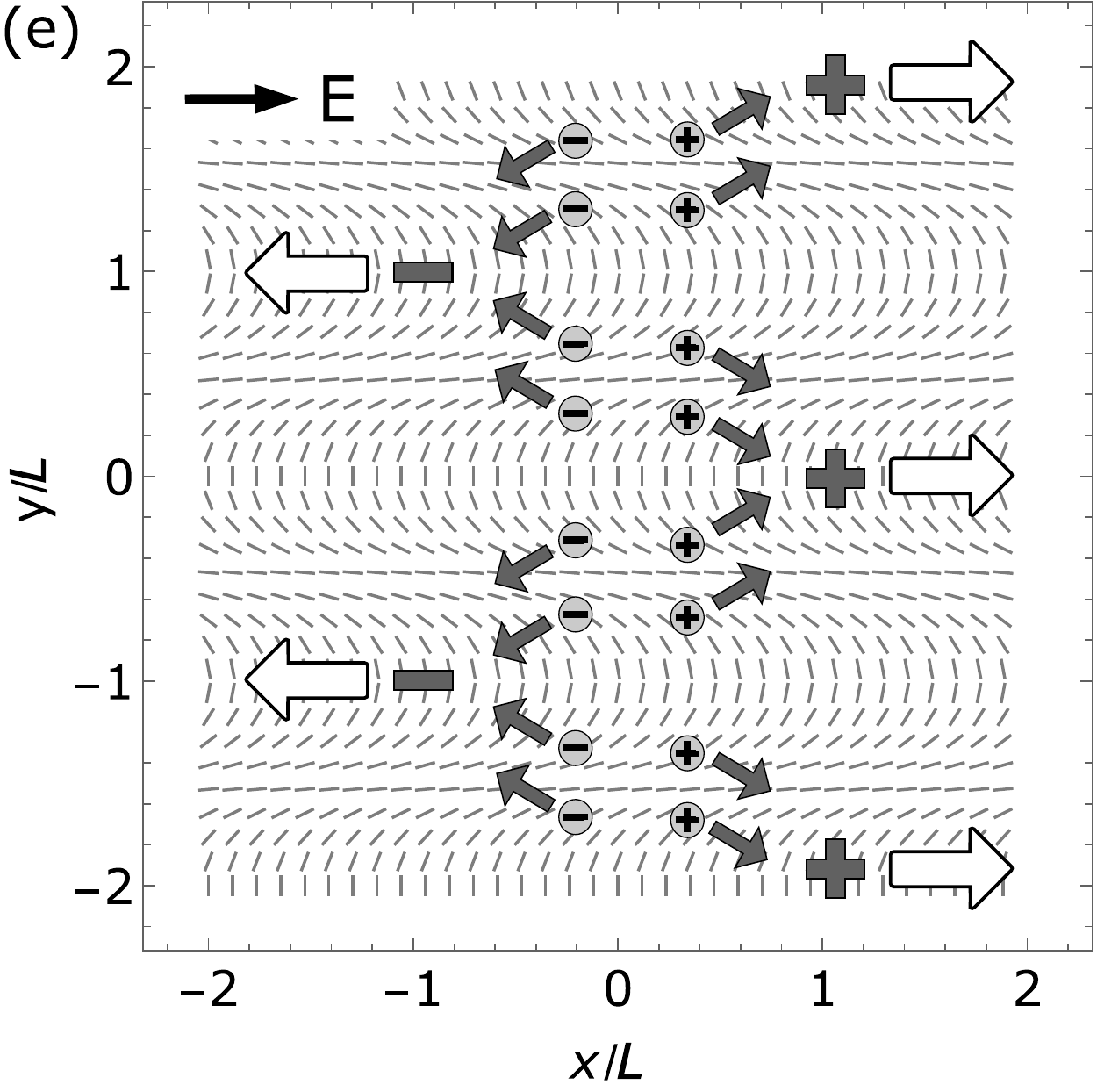}
\includegraphics[width=.3\textwidth]{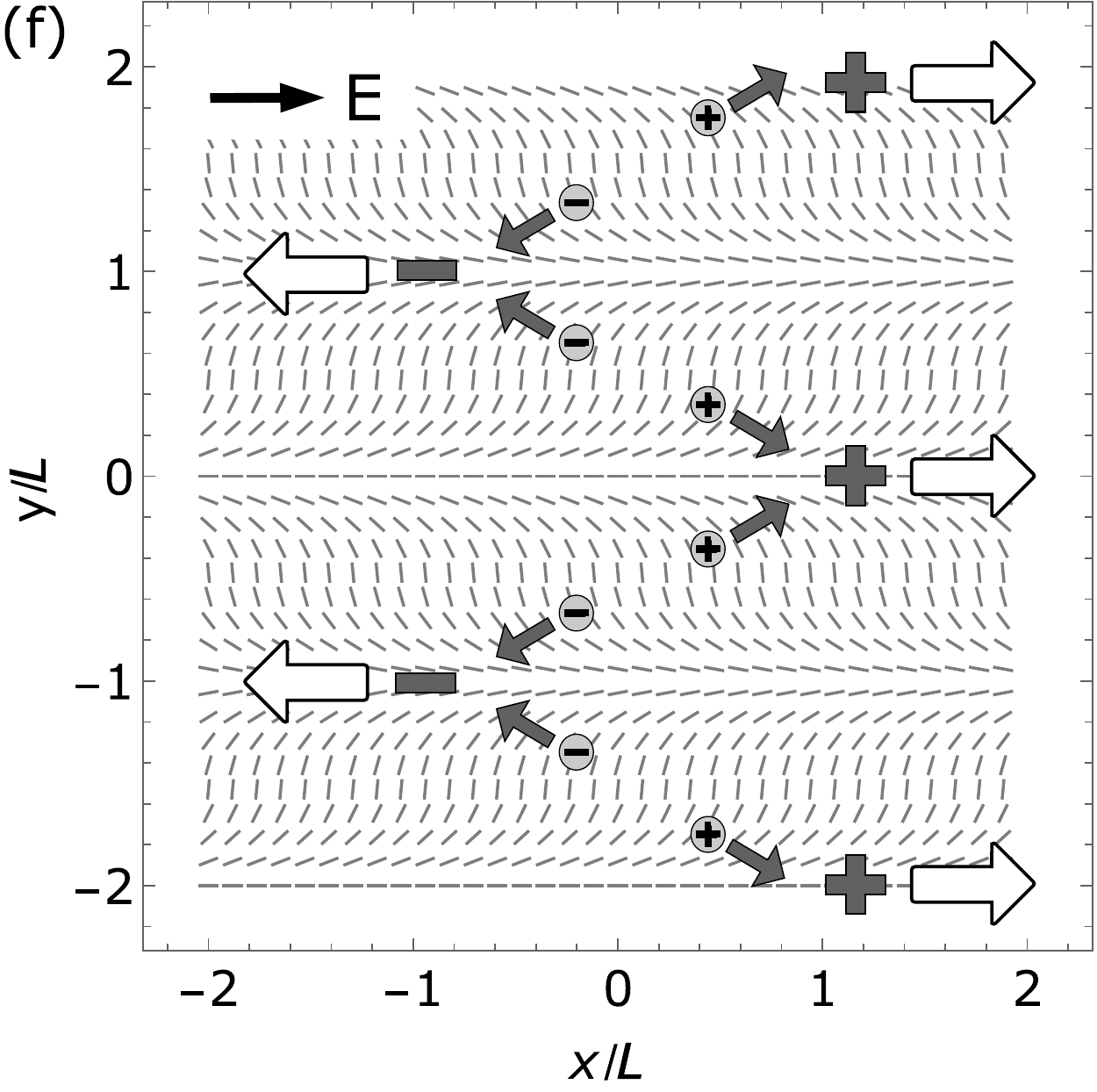}
\caption{Sketch of charge separation: (a) and (d) in the cell $A$; (b) and (e) in the cell $B$; (c) and (f) in the cell $C$. White arrows show the direction of the triggered electro-osmotic flow of the liquid crystal. The key parameters $\lambda_{\sigma}=\sigma_{\|}/\sigma_{\perp}$ and $\lambda_{\varepsilon}=\varepsilon_{\|}/\varepsilon_{\perp}$ are defined as the ratios of the nematic conductivity and permittivity along and perpendicular to the director, respectively. Here (a)-(c): $\lambda_{\varepsilon}-\lambda_{\sigma}<0$; (d)-(f): $\lambda_{\varepsilon}-\lambda_{\sigma}>0$.}\label{Patterns}
\end{center}
\end{figure}

Next we employ the approach developed above to model electro-osmosis in a nematic electrolyte with a prescribed and fixed director distribution.
Such a problem corresponds to recent experiments \cite{Peng_pattern} where a nematic was confined between two parallel substrates with patterned planar anchoring conditions (see Fig.~\ref{Patterns}).
In the presence of anisotropy of dielectric permittivity and/or conductivity, an in-plane electric field would cause director realignment.  The corresponding distortions of the director, however, may be essentially suppressed if the anchoring on the boundary with the substrates is strong enough.  The experiments in Ref.~\cite{Peng_pattern} show that the realignment is indeed small, but this does not indicate that the anchoring is strong since the material studied in Ref.~\cite{Peng_pattern} had zero dielectric anisotropy, $\Delta\varepsilon =0$. 

In order to conclude that the director is essentially unaffected by the field, we need to estimate the minimum value, $W_{\text{min}}$, of the in-plane anchoring strength $W$ that is sufficient to resist the dielectric realignment torque  $\sqrt{\varepsilon_0\Delta\varepsilon K}|E|\delta\varphi$.
Balancing the dielectric realignment torque with the stabilizing surface torque,  $W\delta\varphi$, yields the expression $W_{\text{min}} = \sqrt{\varepsilon_0\Delta\varepsilon K}|E|$. Here $\delta\varphi$  is the small angle of deviation from the anchoring-imposed local director orientation.  

For the typical  parameter values, $\varepsilon_0\Delta\varepsilon =10^{-11}$ F/m, $K=10^{-11}$ N and  $E = 4\times 10^4$ V/m, one finds that $W_{\text{min}}=4\times 10^{-7}$ N/m. 
The actual anchoring strength in the experiments \cite{Peng_pattern} is higher, on the order of  $W\approx K/l \approx 2\times 10^{-6}$ N/m, where $1/l\approx 0.2\times 10^6$ m$^{-1}$ is the highest value of the director gradient that the surface anchoring can support in the experiments \cite{Peng_pattern}. We thus conclude that the director distortions caused by the dielectric torque of the applied electric field can be neglected. Moreover, since the Ericksen number $Er=\alpha_4vL/K$, the ratio of viscous and elastic forces in the experiments \cite{Peng_pattern} is of the order $O(1)$ because $\alpha_4=0.08$ Pa$\cdot$s, $v=4$ $\mu$m/s and $L=50$ $\mu$m (stripes width in Fig.~\ref{Patterns}). Thus we may consider the director field to be ``frozen'' in the first approximation, i.e., it is entirely specified by the surface pattern of molecular orientation $\mathbf{n}=(n_x, n_y)$. Note that the length-scale $l$ above is the minimum distance over which the director gradients are sustained by the surface alignment; it is always much smaller than the typical period of director distortions $L$ in the plane of the liquid crystal cell. 

In the presence of an electric field, spatial variations of $\mathbf{n}(\mathbf{r})$ along with the anisotropy of dielectric permittivity and mobilities of the ions give rise to the separation of electric charges, that are always present in practice.
This field-induced charge density, which is proportional to the field strength $E$, consequently causes a flow of the liquid crystal with the velocity $\propto E^2$.
As we already mentioned above, this remarkable feature allows the flow to be triggered by an alternating current field.   
However, in order to simplify our analysis, we assume that the frequency of the field is much lower than the inverse relaxation time of the ionic gas.
Then we can treat the field as time-independent and content ourselves with steady flows of the liquid crystal. 

Following the experimental setup, specify the director components $n_x=\cos\theta(y)$ and $n_y=\sin\theta(y)$ as periodic functions of $y$, and subject the system to a uniform electric field $\mathbf{E}=(E,0)$ (see Fig.~\ref{Patterns}).
For simplicity, we assume that there are only two ionic species with $z^+=1$ and $z^-=-1$ and concentrations $c^+$ and $c^-$, respectively.

Since the physical system is invariant under arbitrarily translations along the $x$ axis, we seek solutions to the equations \eqref{The_System} in the form
\begin{equation}
\mathbf{v}=(v_x(y), 0),\qquad c^{\pm}=c^{\pm}(y),\qquad \Phi(x, y) = -Ex+\phi(y).
\end{equation} 
Due to incompressibility of the liquid crystal, $\nabla\cdot\mathbf{v}=\partial_x v_x+\partial_y v_y =0$, the velocity component $v_y$ has to be constant and thus can be set to zero. 
Such a velocity field results in vanishing convective derivatives $\mathbf{v}\cdot\nabla$ of $c^{\pm}$, $\mathbf{n}$ and $\mathbf{v}$.
The latter is negligible also in the case of more complex patterns with similar characteristics because of the low Reynolds number $Re = \rho vL/\alpha_4 \approx 2.5\times 10^{-6}$, where $\rho=1$ g/cm$^3$ is the typical liquid crystal density.

Usually, the mobilities of positive and negative ions in nematic liquid crystals are quite close.
Therefore, we can set $\mathsf{D}_{ij}^{+}=\mathsf{D}_{ij}^{-}=\mathsf{D}_{ij}=\bar{\mathsf{D}}\left(\delta_{ij}+(\lambda_{\sigma}-1) n_i n_j\right)$, where $\bar{\mathsf{D}}>0$ and $\lambda_{\sigma} \geq 0$.
The dimensionless parameter $\lambda_{\sigma}=\sigma_{\|}/\sigma_{\perp}$ defined as the ratio of the conductivity (ionic mobility) respectively along and perpendicular to the director, characterizes the anisotropy of the liquid crystal electrolyte.
Under these conditions, the system of governing equations \eqref{The_System} reads as
\begin{equation}\label{The_System_dim_full}
\begin{cases}
-\partial_y \left[\frac{c^{\pm}}{k_B\Theta} \left(\mathsf{D}_{xy}(\partial_x \mu^{\pm}) +\mathsf{D}_{yy}(\partial_y \mu^{\pm})\right) \right] = 0,\\
\partial_x p +\partial_y \left[ \frac{\partial \mathcal{E}_{OF}}{\partial(\partial_y n_x)}(\partial_x n_x) +\frac{\partial \mathcal{E}_{OF}}{\partial(\partial_y n_y)}(\partial_x n_y) -\mathsf{T}_{xy}^V \right] +c^{+}(\partial_x \mu^{+}) +c^{-}(\partial_x \mu^{-}) =0,\\
\partial_y \left[ \varepsilon_{\perp}E_y  +\Delta\varepsilon n_y \left(n_x E_x+n_y E_y\right)\right] = \frac{e}{\varepsilon_0}\left(c^{+}-c^{-}\right),\\
\partial_y \left[ \frac{\partial \mathcal{E}_{OF}}{\partial(\partial_y n_x)}(\partial_y n_x) +\frac{\partial \mathcal{E}_{OF}}{\partial(\partial_y n_y)}(\partial_y n_y) +p -\mathsf{T}_{yy}^V \right] +c^{+}(\partial_y \mu^{+}) +c^{-}(\partial_y \mu^{-})+\hfill\\ 
\hfill +\varepsilon_0\Delta\varepsilon \left(n_x E_x +n_y E_y\right)\left( E_x(\partial_y n_x) +E_y(\partial_y n_y)\right)=0.\\
\end{cases}
\end{equation}
Note that although in the original experiments \cite{Peng_pattern} the liquid crystal was dielectrically isotropic, $\Delta\varepsilon =0$, here we keep the terms with $\Delta\varepsilon$ in order to explore a role of this sort of anisotropy as well. 

Within the commonly adopted one-constant approximation $K_1=K_2=K_3=K$ and $\mathcal{E}_{OF}=\frac{1}{2}K(\partial_i n_j)(\partial_i n_j)$.
Thus, 
\begin{eqnarray}
\frac{\partial \mathcal{E}_{OF}}{\partial(\partial_y n_x)}(\partial_x n_x) +\frac{\partial \mathcal{E}_{OF}}{\partial(\partial_y n_y)}(\partial_x n_y) =0,\\
\frac{\partial \mathcal{E}_{OF}}{\partial(\partial_y n_x)}(\partial_y n_x) +\frac{\partial \mathcal{E}_{OF}}{\partial(\partial_y n_y)}(\partial_y n_y) =K\left(\frac{d\theta}{dy}\right)^2.
\end{eqnarray}
The viscous stress tensor \eqref{Viscous_stress} for the system under investigation reduces to $\mathsf{T}_{xy}^V=\alpha_4 \eta(y)\partial_y v_x$ and $\mathsf{T}_{yy}^V=\alpha_4\chi(y)\partial_y v_x$, where the functions $\eta(y)$ and $\chi(y)$ are defined by the nematic's viscosities and its director field,
\begin{eqnarray}
\eta(y) = \frac{1}{2} +\frac{\alpha_3+\alpha_6}{2\alpha_4}n_x^2 +\frac{\alpha_5-\alpha_2}{2\alpha_4}n_y^2 +\frac{\alpha_1}{\alpha_4}n_x^2n_y^2,\\
\chi(y) = \frac{\alpha_1}{\alpha_4}n_xn_y^3 +\frac{\alpha_6}{\alpha_4}n_xn_y.
\end{eqnarray}
Interestingly, $\eta(y)$ can be easily expressed in terms of the Miesowicz viscosities \cite{Mie} defined as $\eta_1 = \frac{1}{2}(\alpha_3+\alpha_6+\alpha_4)$ measured when $\mathbf{n}$ is parallel to $\mathbf{v}$, $\eta_2 = \frac{1}{2}(\alpha_5-\alpha_2+\alpha_4)$ measured when $\mathbf{n}$ is parallel to $\nabla\mathbf{v}$, $\eta_3 = \frac{1}{2}\alpha_4$ measured with $\mathbf{n}$ orthogonal to both $\mathbf{v}$ and $\nabla\mathbf{v}$, and $\eta_{12}=\alpha_1$ so that
\begin{equation}
\eta(y) = \frac{1}{2}\left(\frac{\eta_1}{\eta_3}n_x^2 +\frac{\eta_2}{\eta_3} n_y^2 +\frac{\eta_{12}}{\eta_3}n_x^2n_y^2\right).
\end{equation}
Taking a closer look at the last equation in \eqref{The_System_dim_full}, we observe that it gives the pressure $p$ as a function of the $y$-coordinate, whereas the remaining part of the system includes only $\partial_x p$.
Hence, the latter is zero unless the external pressure gradient is applied to the system.
As a result, $p(y)$ can be found when the reduced system
\begin{equation}\label{The_System_dim}
\begin{cases}
-\partial_y \left[\frac{c^{\pm}}{k_B\Theta} \left(\mathsf{D}_{xy}(\partial_x \mu^{\pm}) +\mathsf{D}_{yy}(\partial_y \mu^{\pm})\right) \right] = 0,\\
-\partial_y\mathsf{T}_{xy}^V +c^{+}(\partial_x \mu^{+}) +c^{-}(\partial_x \mu^{-}) =0,\\
\partial_y \left[ \varepsilon_{\perp}E_y  +\Delta\varepsilon n_y \left(n_x E_x+n_y E_y\right)\right] = \frac{e}{\varepsilon_0}\left(c^{+}-c^{-}\right).
\end{cases}
\end{equation}
is solved.  

It is convenient to nondimensionalize the problem \eqref{The_System_dim} by introducing new variables
\begin{equation}
\tilde{y} = \frac{y}{L}, \quad \tilde{c}^{\pm}=\frac{c^{ \pm}}{\bar{c}}, \quad \tilde{\phi}=\frac{\phi}{EL}, \quad \tilde{\mathsf{D}}_{ij}=\frac{\mathsf{D}_{ij}}{\bar{\mathsf{D}}},
\end{equation}  
where $L$ denotes the stripe's width for a given pattern (see Fig.~\ref{Patterns}) and $\bar{c}$ is the average bulk concentration of the ions.
Then, after the tildes are omitted for notational simplicity, the system \eqref{The_System_dim} reads as  
\begin{equation}\label{The_System_nondim}
\begin{cases}
\partial_y \left[F\mathsf{D}_{xy}(-c^{+})+\mathsf{D}_{yy}\left(\partial_y c^{+} +Fc^{+}(\partial_y\phi)\right) \right] = 0,\\
\partial_y \left[F\mathsf{D}_{xy}(c^{-})+\mathsf{D}_{yy}\left(\partial_y c^{-} -Fc^{-}(\partial_y\phi)\right) \right] = 0,\\
\partial_y \left[ -\partial_y\phi  +\Delta_{\varepsilon} n_xn_y -\Delta_{\varepsilon} n_y^2(\partial_y\phi)\right] = G\left(c^{+}-c^{-}\right),\\
\partial_y \left[\eta(y) \partial_y v_x\right] =-G(c^{+}-c^{-}),\\
\end{cases}
\end{equation}
where nondimensional parameters
\begin{equation}
\Delta_{\varepsilon}=\frac{\varepsilon_{\|}-\varepsilon_{\perp}}{\varepsilon_{\perp}}=\lambda_{\varepsilon}-1,\qquad
G=\frac{\bar{c}eL}{\varepsilon_0\varepsilon_{\perp}E},\qquad 
F=\frac{eEL}{k_{B}\Theta}
\end{equation}
and a characteristic value for the velocity quadratic in the field strength, $\bar{v}=\varepsilon_{\perp}\varepsilon_{0}LE^2/\alpha_4$, emerges naturally.

Consider the first two equations in \eqref{The_System_nondim}.
It follows that
\begin{equation}\label{NP_nondim}
\begin{split}
F\mathsf{D}_{xy}(-c^{+})+\mathsf{D}_{yy}\left(\partial_y c^{+} +Fc^{+}(\partial_y\phi)\right)= \text{const}_1, \\
F\mathsf{D}_{xy}(c^{-})+\mathsf{D}_{yy}\left(\partial_y c^{-} -Fc^{-}(\partial_y\phi)\right)= \text{const}_2.
\end{split}
\end{equation}
In order to find the unknown constants we have to recall that the left-hand sides of \eqref{NP_nondim} define the flux of corresponding ions along the $y$ axis, $J_y^{\pm}\propto c^{\pm}\mathsf{D}_{yi}(\partial_i\mu^{\pm})$.
Since there is no reason for such a uniform constant flux to exist, both $\text{const}_1$ and $\text{const}_2$ have to vanish.
Dividing the first and the second equations \eqref{NP_nondim} by $c^+$ and $c^-$, respectively, and adding the results, one easily arrives at $c^+c^-=c_0^2$, where $c_0$ is a non-zero constant.
Given this fact, the subtraction of the equations \eqref{NP_nondim} leads to
\begin{equation}\label{E_y}
\partial_y\phi = \frac{\mathsf{D}_{xy}}{\mathsf{D}_{yy}} -\frac{1}{F}\partial_y\left(\ln c^+\right).
\end{equation}
Let $\ln c^+=r$ and substitute \eqref{E_y} into the third equation of the system \eqref{The_System_nondim},
\begin{equation}\label{Eq_for_r}
\frac{1}{F}\left(r^{\prime} \left(1+\Delta_{\varepsilon}n_y^2 \right) \right)^{\prime} -G\left( e^r -c_0^2e^{-r} \right) =\left( \frac{\mathsf{D}_{xy}}{\mathsf{D}_{yy}}\left(1+\Delta_{\varepsilon}n_y^2\right) -\Delta_{\varepsilon}n_x n_y \right)^{\prime}.
\end{equation}
Hereafter the prime denotes total derivative $\frac{d}{dy}$.
The number of ions present in the system is fixed, therefore
\begin{equation}\label{Cond_for_r}
\int_{-N}^{N}e^rdy=c_0^2 \int_{-N}^{N}e^{-r}dy=2N.
\end{equation}
Equation \eqref{Eq_for_r} and the condition \eqref{Cond_for_r} make it possible to find the concentrations of both ionic species, which, in turn, allow us to calculate all the remaining unknown quantities. The problem \eqref{Eq_for_r}-\eqref{Cond_for_r} is essentially nonlinear and its analytical solutions are difficult to find in the general case (see Figs.~\ref{Numer_good} -- \ref{Numer_diffr} for numerical solutions).

However, for the parameters characterizing the experimental setup in \cite{Peng_pattern}, we have $E = 40$~mV$/\mu$m, $\bar{c} = 10^{19}$~m$^{-3}$, $L=50$~$\mu$m and $\varepsilon_{\perp}=6$  (the values that are quite typical for nematic systems), and the Eq.~\eqref{Eq_for_r} can be linearized provided that the liquid crystal is not strongly dielectrically anisotropic, $\Delta_{\varepsilon}\sim 1$.  

But first note that the electric field applied parallel to the sandwich-like cells with a thin nematic layer confined between two glass plates is, generally speaking, spatially nonuniform \cite{Lazo_2}.
In particular, its value is diminished in the center of the cell;  the reduction factor for experimental conditions close to the ones discussed in this paper is about 0.6 \cite{Lazo_2}, thus the applied electric field $40$~mV$/\mu$m  is reduced to about $24$~mV$/\mu$m.

For the listed values, $F\approx 47$ and $G\approx 63$ are sufficiently large so that $\delta=\frac{1}{F}$ can be treated as a small parameter, $\delta\ll 1$. 
Then \eqref{Eq_for_r} takes the form
\begin{equation}\label{Eq_for_r_app}
\delta\left(r^{\prime} \left(1+\Delta_{\varepsilon}n_y^2 \right) \right)^{\prime} -\frac{b}{\delta}\left( e^r -c_0^2e^{-r} \right) =-M^{\prime}(y),
\end{equation}
where the right-hand side is denoted as $M^{\prime}(y)$ with
\begin{equation}\label{M}
M(y) = -\frac{(\lambda_{\sigma}-1)n_xn_y}{1+(\lambda_{\sigma}-1)n_y^2}\left(1+\Delta_{\varepsilon}n_y^2\right) +\Delta_{\varepsilon}n_xn_y =(\lambda_{\varepsilon}-\lambda_{\sigma})\frac{n_xn_y}{1+(\lambda_{\sigma}-1)n_y^2}
\end{equation}
and $b=G{\delta}=\mathcal{O}(1)$.
Taking into account smallness of $\delta$, we can approximate $r$ and $c_0$ by  
\begin{equation}\label{Linearization}
r=\delta r_1 +\mathcal{O}(\delta^2)\quad\text{and}\quad c_0=1+c_1\delta +\mathcal{O}(\delta^2),
\end{equation}
which implies that deviations of the concentrations $c^{\pm}$ from the average value $\bar{c}$ are small.
Thus, to leading order in $\delta$ equations \eqref{Eq_for_r_app} and \eqref{Cond_for_r} result in
\begin{equation}
r_1 -c_1 = \frac{1}{2b}M^{\prime}(y)
\end{equation}
and
\begin{equation}
\label{eq:ingr}
\int_{-N}^{N}\left( 1 +\delta r_1 \right)dy = \int_{-N}^{N}\left( 1 -\delta r_1 \right)dy +2c_1\delta\int_{-N}^{N}dy =2N,  
\end{equation}
respectively. 
Hence, $c_1=0$ and
\begin{equation}\label{Concentration}
c^{\pm}=1 \pm\delta r_1 = 1 \pm\frac{1}{2G}M^{\prime}(y),
\end{equation}
which clearly shows that the (nondimensional) electric charge density $Q(y)=c^{+}-c^{-}$ is indeed proportional to the field strength as $1/G\propto E$.

Finally, consider the last equation of the system \eqref{The_System_nondim}
\begin{equation}
\left( \eta v_x^{\prime} \right)^{\prime} = -G\left(c^{+}-c^{-}\right) =-M^{\prime}
\end{equation}
which defines the velocity of the flow.
It is a second-order differential equation so that its general solution
\begin{equation}\label{Velocity_gen}
v_x = -\int dy\frac{M(y)+C_1}{\eta(y)} +C_2,
\end{equation}
contains two unknown constants $C_1$ and $C_2$. In order to find these constants, we have to specify the director field as a function of $y$, i.e., the explicit form of $\theta=\theta(y)$.

\subsection{Pattern $A$}

\begin{figure}
\begin{center}
\includegraphics[width=.3\textwidth]{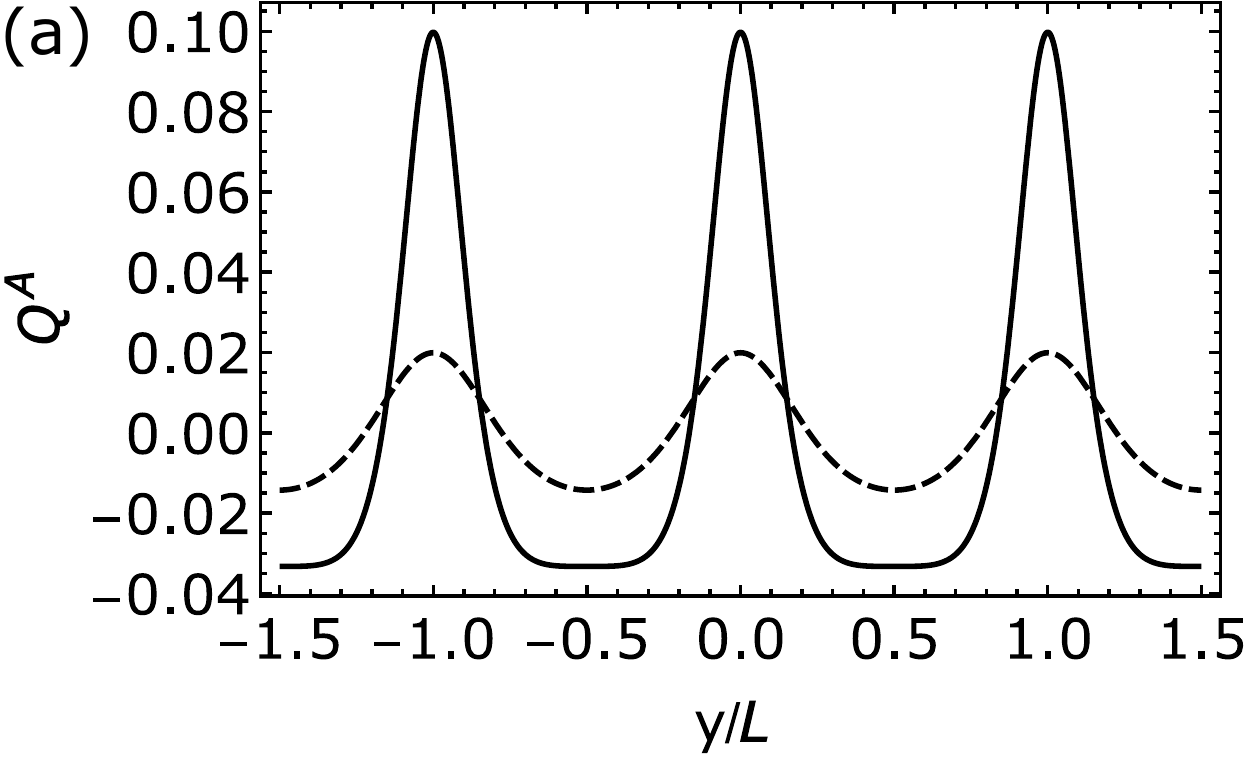}
\includegraphics[width=.3\textwidth]{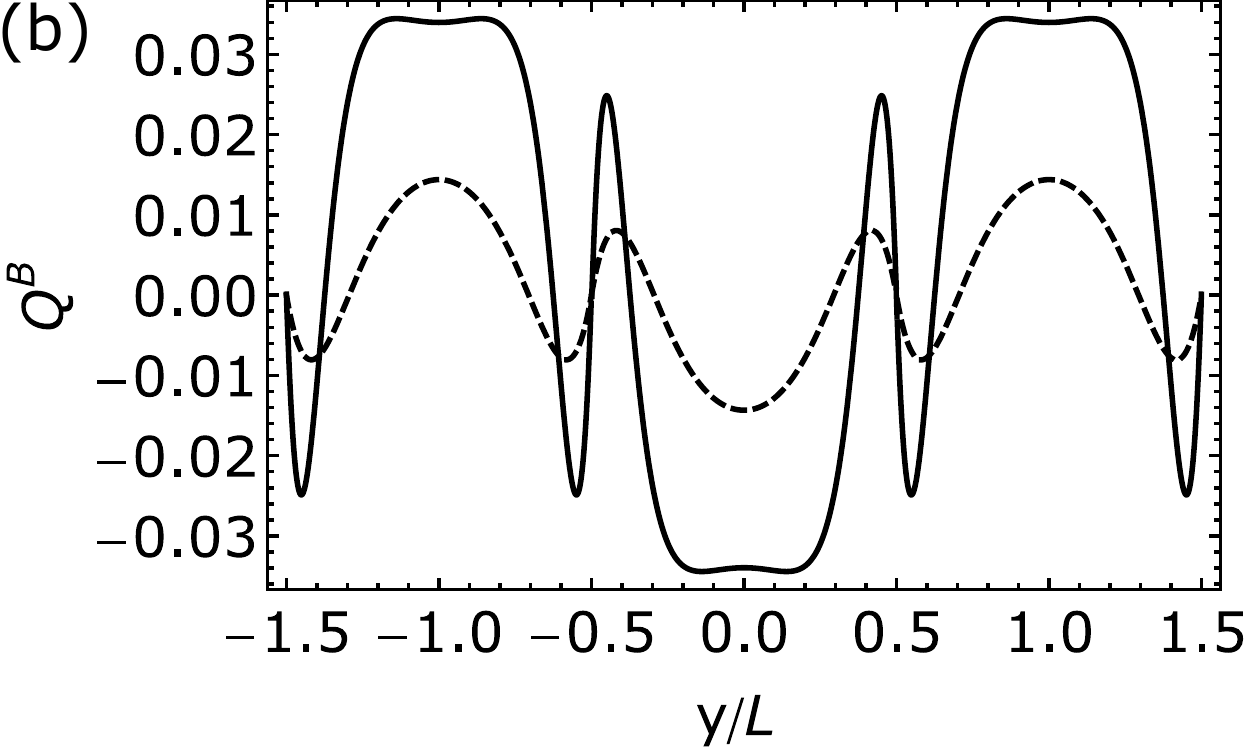}
\includegraphics[width=.3\textwidth]{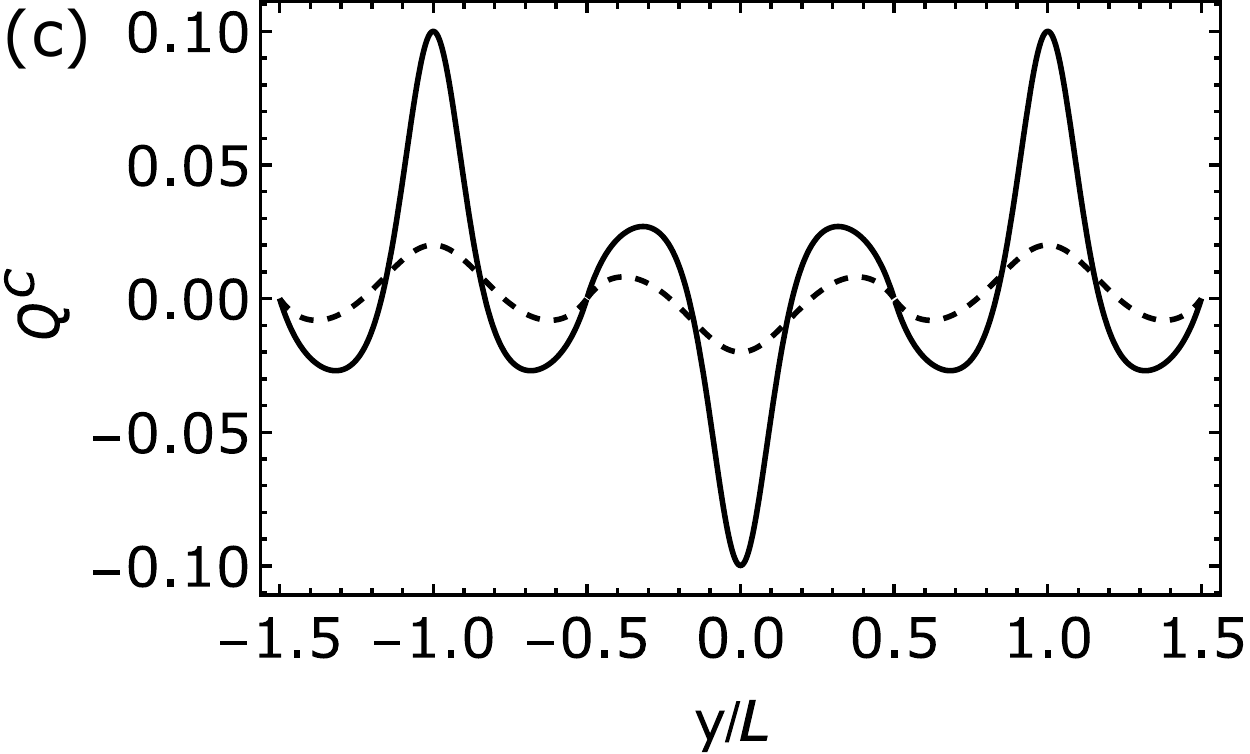}\\
\includegraphics[width=.3\textwidth]{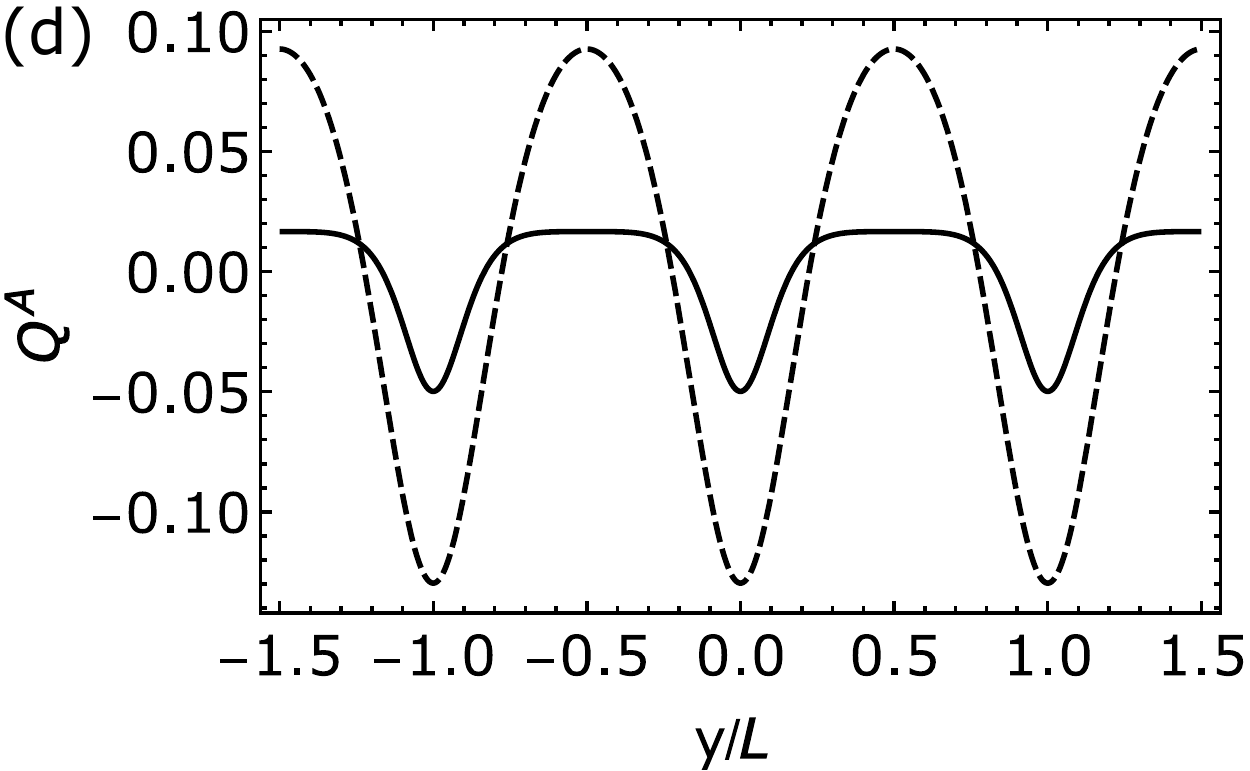}
\includegraphics[width=.3\textwidth]{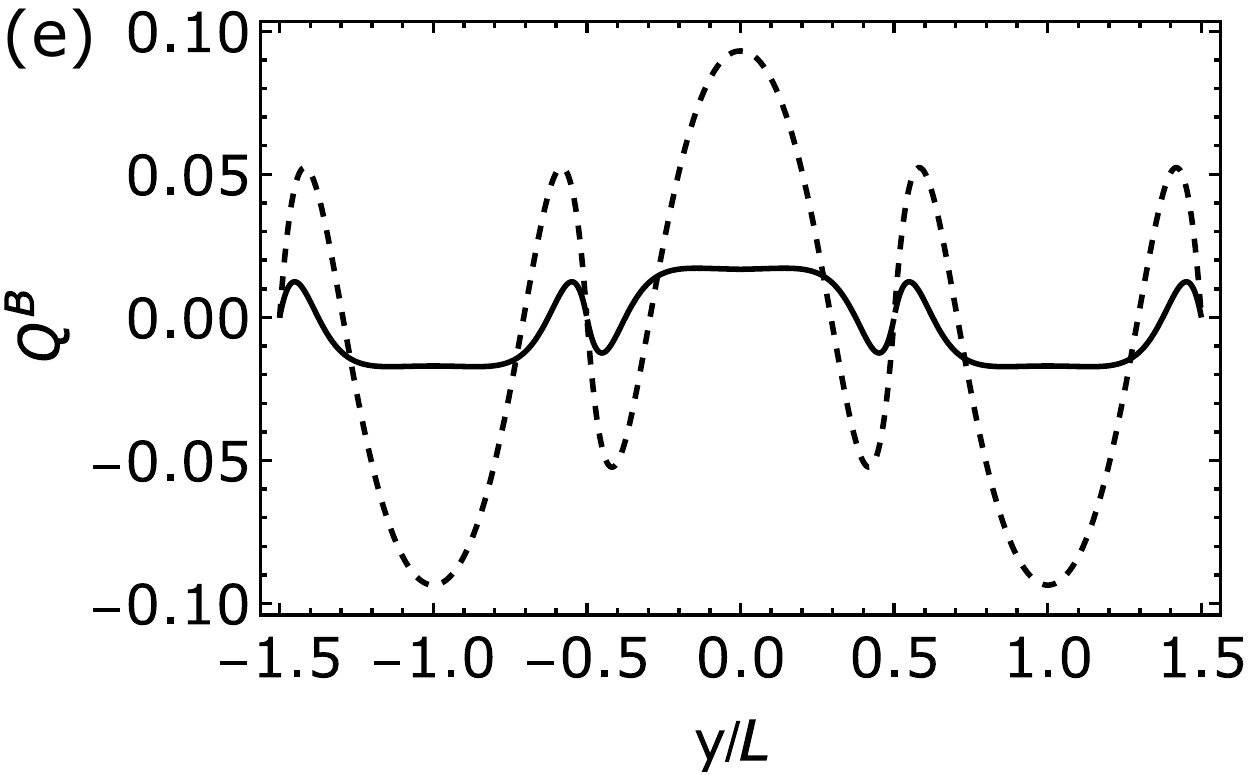}
\includegraphics[width=.3\textwidth]{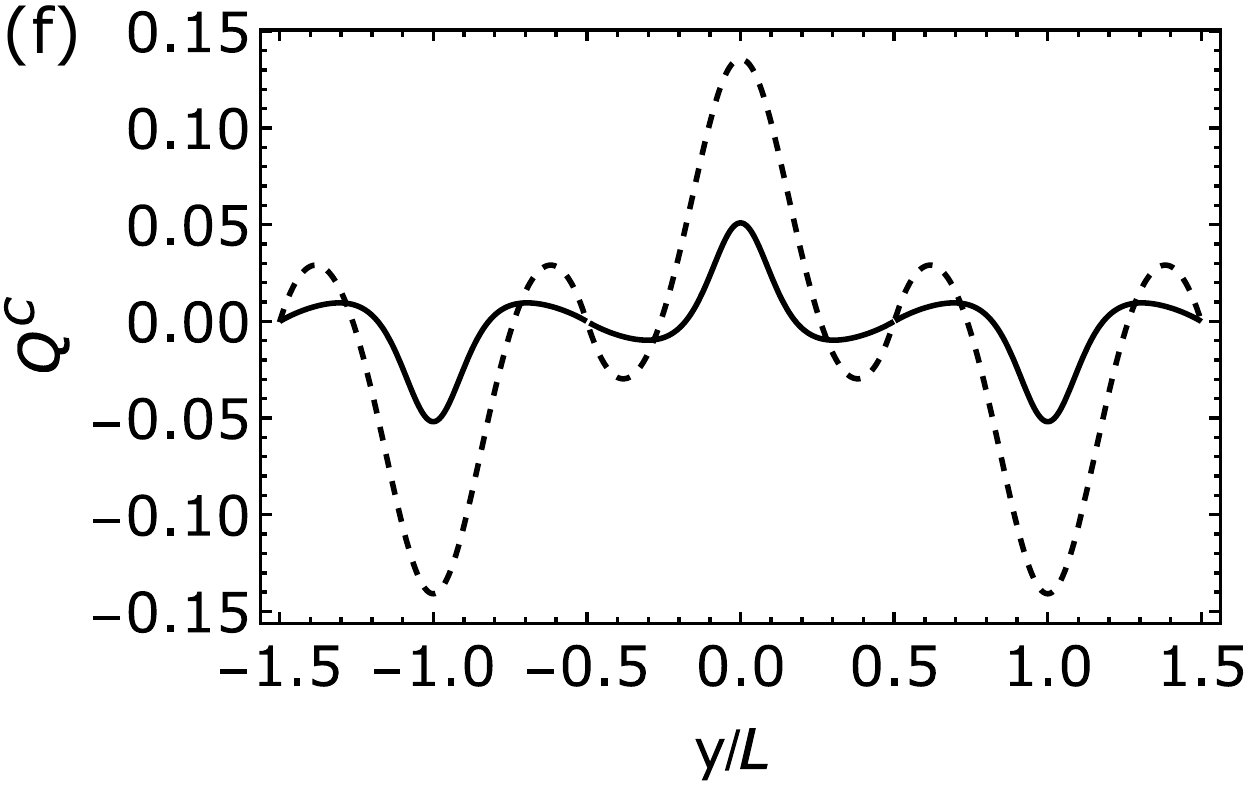}
\caption{Nondimensional charge concentration: (a) and (d) in the cell $A$; (b) and (e) in the cell $B$; (c) and (f) in the cell $C$. Here
(a)-(c): $\lambda_{\sigma}=1.4$, $\lambda_{\varepsilon}=1$ (dashed line) and $\lambda_{\sigma}=3$, $\lambda_{\varepsilon}=1$ (solid line); 
(d)-(f): $\lambda_{\sigma}=1.4$, $\lambda_{\varepsilon}=4$ (dashed line) and $\lambda_{\sigma}=3$, $\lambda_{\varepsilon}=4$ (solid line).
}\label{Charge_plot}
\end{center}
\end{figure}

Let us start with the pattern given by $\theta^A = \pi(1-y)$, i.e. $\mathbf{n}=(-\cos\pi y,\, \sin\pi y)$.
Suppose, for instance, $\lambda_{\sigma}>1$ so that the mobility of ions along the director is higher than that in the perpendicular direction.
One can roughly think of this as the ions moving mainly along the direction of $\mathbf{n}\equiv -\mathbf{n}$.
Then under the action of the field parallel to the $x$-axis,  the positive ions that are otherwise distributed homogeneously, accumulate in the regions where $\mathbf{n}=(1, 0)$ whereas the negative ions migrate to the regions where $\mathbf{n}=(0, 1)$.
At the same time, positive dielectric anisotropy, $\lambda_{\varepsilon}>1$ or $\Delta\varepsilon >0$, gives rise to the opposite pattern of charge separation.
Response of the liquid crystal to the electric field requires an excess of positive and negative charges in the regions with $\mathbf{n}=(0, 1)$ and $\mathbf{n}=(1, 0)$, respectively.
Equations \eqref{Concentration} and \eqref{M} transparently reflect this interplay between the two mechanisms through the multiplier $\lambda_{\varepsilon}-\lambda_{\sigma}$.
In particular, the electric charge distribution $Q^A=c^{+}-c^{-}$ in the $A$-type cell is given by
\begin{equation}\label{Concentartion_A}
Q^A = \frac{\pi(\lambda_{\varepsilon}-\lambda_{\sigma})}{2G}\,\frac{\lambda_{\sigma}-1-(1+\lambda_{\sigma})\cos 2\pi y}{\left( 1+(\lambda_{\sigma}-1)\sin^2\pi y\right)^2}.
\end{equation} 
It follows from \eqref{Concentartion_A} that the dielectric anisotropy by itself is capable of separating the ions.
It should be noted, however, that $\lambda_{\varepsilon}$ and $\lambda_{\sigma}$ are not interchangeable.
The charge distribution $Q^A$ is linear in $\lambda_{\varepsilon}$, while it depends on $\lambda_{\sigma}$ in a more complex way.
If there is no difference between the mobilities, $\lambda_{\sigma}=1$, the charges are symmetrically separated, $Q^A = -\pi(\lambda_{\varepsilon}-1)G^{-1}\cos 2\pi y$.
Equal amounts of positively or negatively charged ions are distributed over equal amounts of the liquid crystal; otherwise the symmetry between cations and anions is broken (see Fig.~\ref{Charge_plot}).
When the mobilities in the directions along and perpendicular to $\mathbf{n}$ differ considerably, a certain number of ions is practically trapped within the regions of low mobility.
Only the ``fast'' charges separate under this condition.
As a result, we see narrow peaks of either positive or negative charge, depending on the sign of $\lambda_{\varepsilon}-\lambda_{\sigma}$, separated by wide plateaus of the opposite charge in Fig.~\ref{Charge_plot}.

Once the charges have separated, their movement causes electrokinetic flow of the liquid crystal with the velocity given by the general expression \eqref{Velocity_gen}.
The director field $\mathbf{n}=(-\cos\pi y,\,\sin\pi y)$ is periodic with the period 1.
Naturally, the same should hold for the velocity.  
The integral $\int dy M/\eta$ results in the function
\begin{multline}\label{Velocity_A}
\int dy M/\eta=\frac{(\lambda_{\varepsilon}-\lambda_{\sigma})}{2\pi}\frac{1}{ (\lambda_{\sigma} -1) \left(\tilde{\eta} _1 \lambda_{\sigma} -\tilde{\eta} _2\right)-\tilde{\eta} _{12} \lambda_{\sigma}}\times\\ 
\times\left\{\frac{\left(\tilde{\eta} _1-\tilde{\eta} _2\right) (\lambda_{\sigma} -1)-\tilde{\eta} _{12} (\lambda_{\sigma} +1)}{\sqrt{\left(\tilde{\eta}_1 +\tilde{\eta} _2+\tilde{\eta} _{12}\right){}^2 -4\tilde{\eta}_{1}\tilde{\eta}_{2}}} \ln \frac{\left|\sqrt{\left(\tilde{\eta}_1-\tilde{\eta}_2 +\tilde{\eta}_{12}\right)^2 +4\tilde{\eta}_{2}\tilde{\eta}_{12}}+\tilde{\eta} _1-\tilde{\eta} _2-\tilde{\eta} _{12} \cos 2 \pi  y\right|}{\left|\sqrt{\left(\tilde{\eta}_1-\tilde{\eta}_2 +\tilde{\eta}_{12}\right)^2 +4\tilde{\eta}_{2}\tilde{\eta}_{12}}-\tilde{\eta} _1+\tilde{\eta} _2+\tilde{\eta} _{12} \cos 2 \pi  y\right|}-\right.\\
\left. -(\lambda_{\sigma} -1) \ln \frac{\left|\left(\tilde{\eta}_1-\tilde{\eta}_2 +\tilde{\eta}_{12}\right)^2 +4\tilde{\eta}_{2}\tilde{\eta}_{12}-\left(\tilde{\eta} _1-\tilde{\eta} _2-\tilde{\eta} _{12} \cos 2 \pi  y\right)^2\right|}{\left(1+\lambda_{\sigma} -(\lambda_{\sigma} -1) \cos 2 \pi  y\right)^2}\right\}
\end{multline}
that is indeed periodic with the period 1.
Here $\tilde{\eta}_1=\eta_1/\eta_3$, $\tilde{\eta}_2=\eta_2/\eta_3$, and $\tilde{\eta}_{12}=\eta_{12}/\eta_3$.
At the same time $\int dy C_1/\eta$ is not periodic.
By definition the viscous function $\eta(y)$ is positive-definite.
Therefore, $C_1/\eta$ does not change its sign.
Due to this fact, the integral $\int dy C_1/\eta$ is a monotonic function of $y$.
If we expect the velocity to be periodic and continuous, $C_1$ has to be zero.
Hence,
\begin{equation}\label{Velocity_A_full}
v_x^A = -\int dy\frac{M(y)}{\eta(y)} +\int_0^1 dy \int dy \frac{M(y)}{\eta(y)},
\end{equation}
where the remaining constant $C_2$ was chosen so as to avoid the net transport of the liquid crystal through the system.

Equations \eqref{Velocity_A} and \eqref{Velocity_A_full} prove that the anisotropy of ionic mobility is not a prerequisite for liquid-crystal-enabled electrokinetics.
Even when $\lambda_{\sigma}=1$ the flow can exist as long as $\lambda_{\varepsilon}\neq 1$.
Its profile, however, will slightly differ from those of the flow caused by the corresponding pair $\lambda_{\sigma}^{\prime}=1/\lambda_{\varepsilon}$, $\lambda_{\varepsilon}^{\prime}=1$ (see Fig.~\ref{Velocity_plot}, bottom row).

The profile of the flow depends also on the  viscosities $\tilde{\eta}_1$, $\tilde{\eta}_2$, and $\tilde{\eta}_{12}$.
This dependence, however, does not lead to any important consequences because the viscosities cannot alter key features of the flow.
For given anisotropies $\lambda_{\varepsilon}$ and $\lambda_{\sigma}$, magnitudes of $\tilde{\eta}_1$, $\tilde{\eta}_2$, and $\tilde{\eta}_{12}$ define amplitudes and zeros of $v_x(y)$.
But they can neither reverse nor distort the flow direction  (at least within the present approach which assumes that $\mathbf{n}(\mathbf{r})$ is fixed). 

Note that in the case $\lambda_{\varepsilon}=1$, the solution \eqref{Velocity_A_full} was already obtained in \cite{Carme} under the initial assumption of isotropic viscosity $\eta(y)\equiv 1/2$.
Our results show that although the assumption may seem oversimplified, it leads to qualitatively correct behavior of the flow (compare dashed and solid lines in the top row of Fig.~\ref{Velocity_plot}). 

\subsection{Patterns $B$ and $C$}

\begin{figure}
\begin{center}
\includegraphics[width=.3\textwidth]{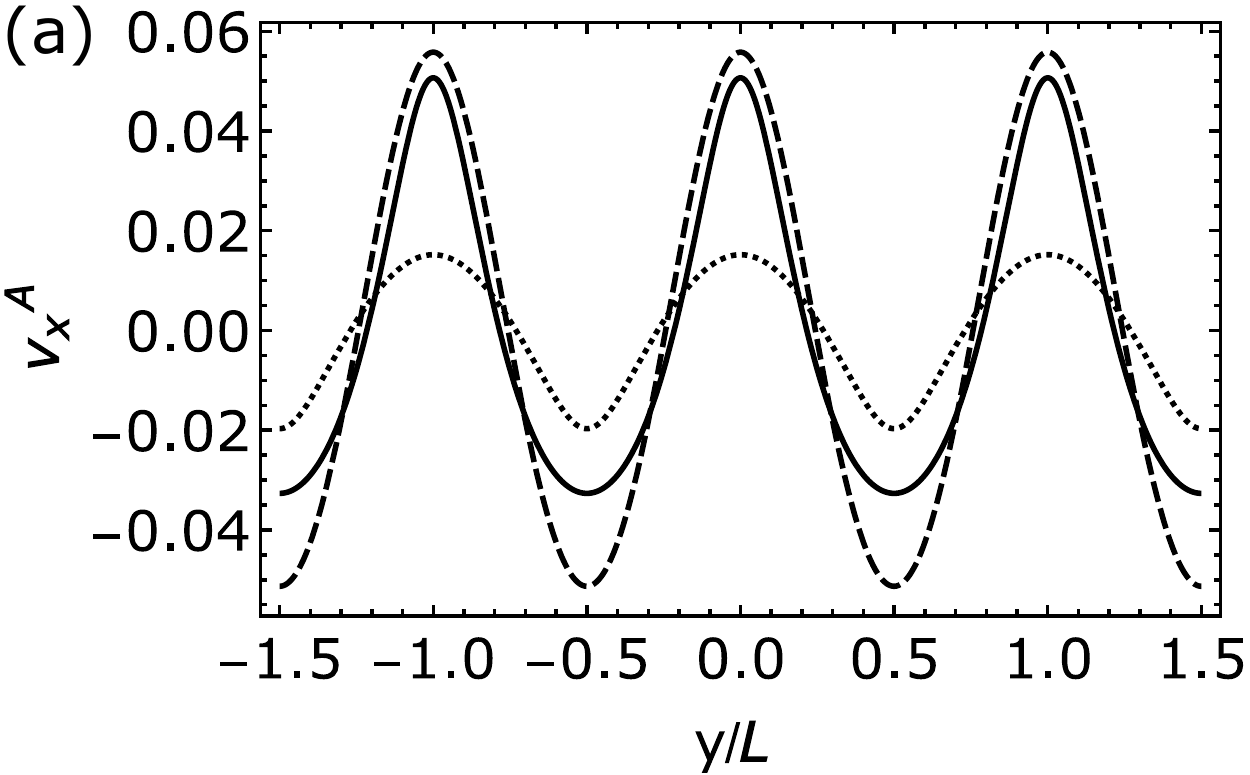}
\includegraphics[width=.3\textwidth]{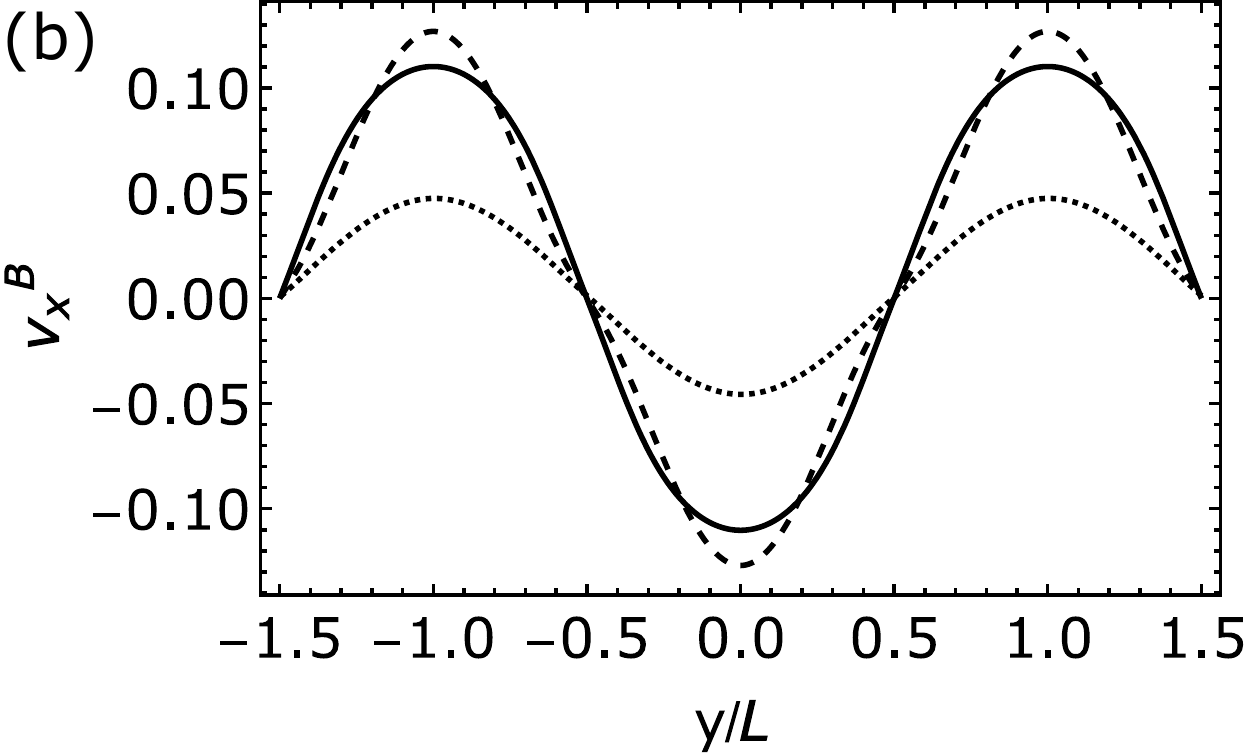}
\includegraphics[width=.3\textwidth]{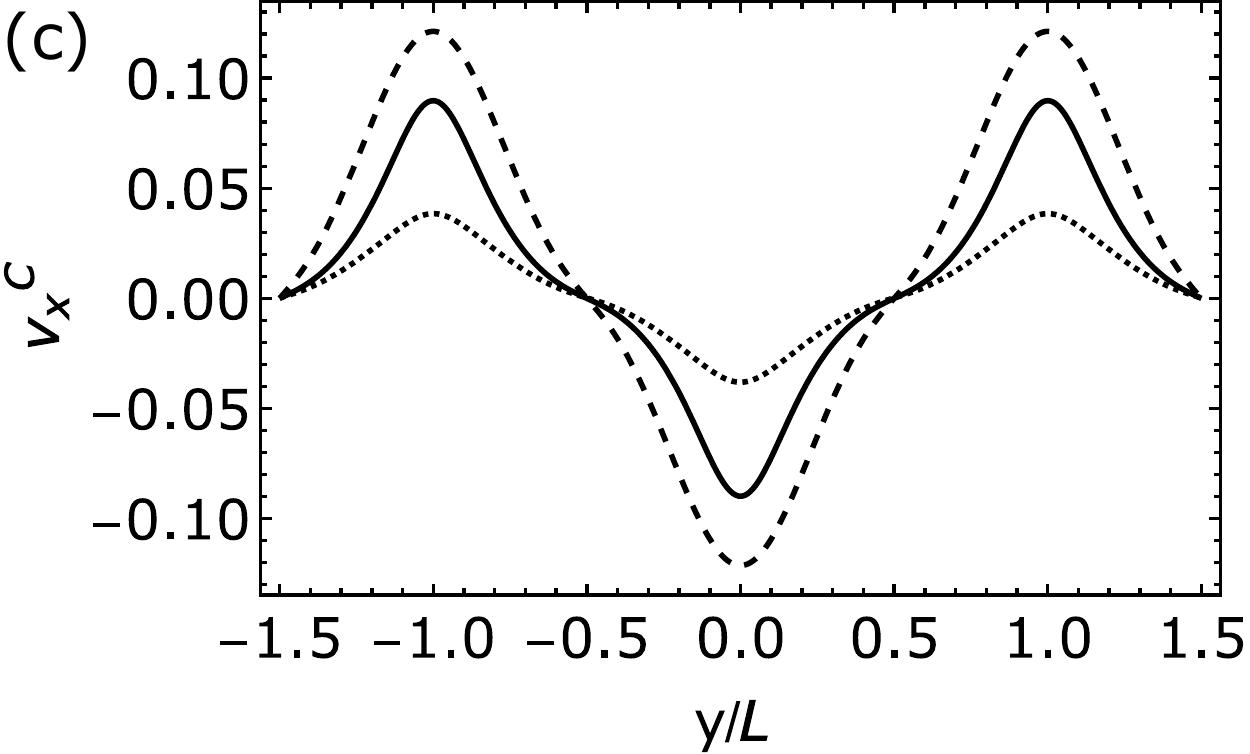}\\
\includegraphics[width=.3\textwidth]{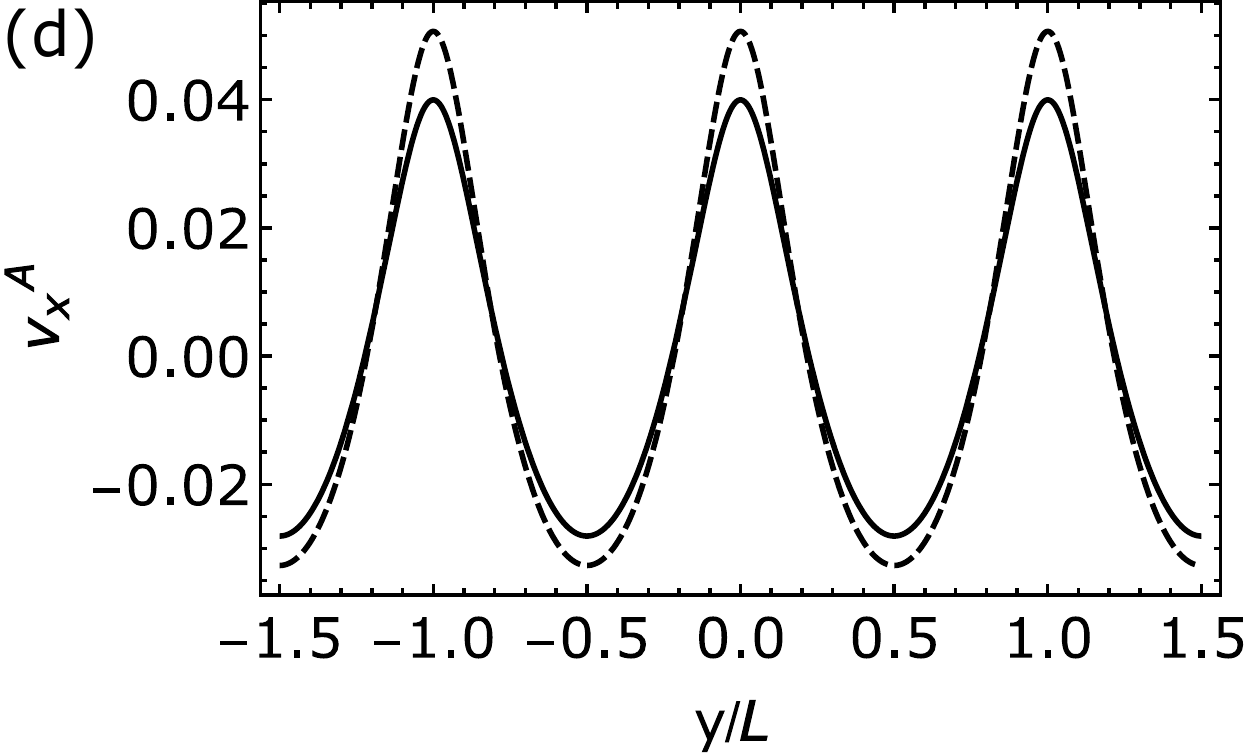}
\includegraphics[width=.3\textwidth]{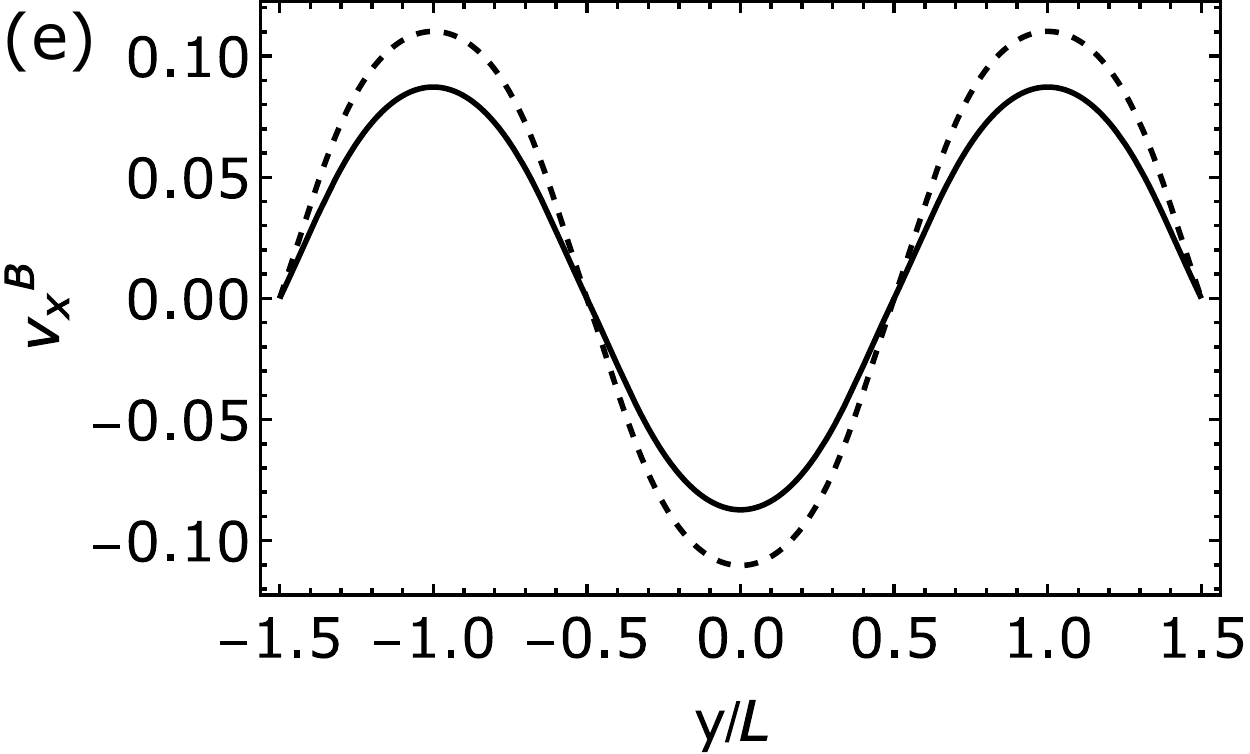}
\includegraphics[width=.3\textwidth]{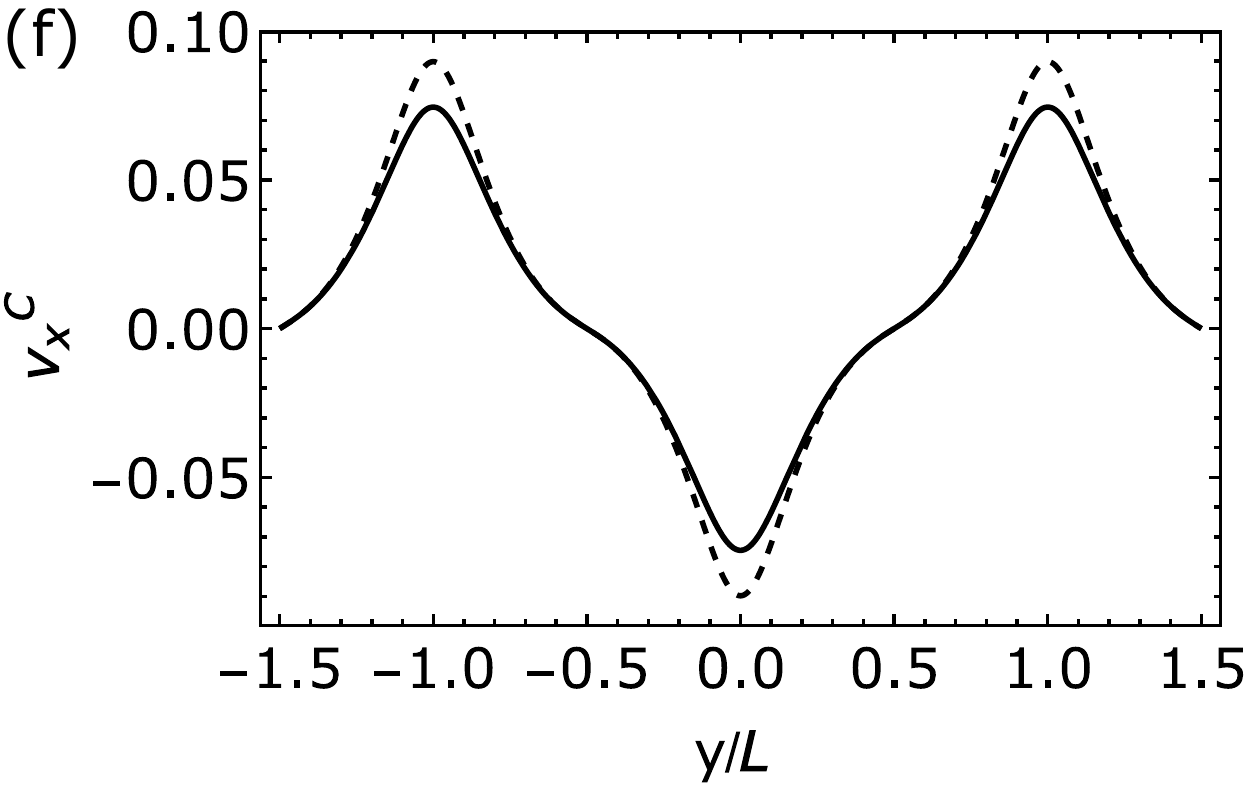}
\caption{Nondimensional flow velocity: (a) and (d) in the cell $A$; (b) and (e) in the cell $B$; (c) and (f) in the cell $C$. Here the plots (a)-(c) correspond to the fixed anisotropy $\lambda_{\varepsilon}=1$, $\lambda_{\sigma}=1.4$ and varying viscosity: $\tilde{\eta}_1=0.6$, $\tilde{\eta}_2=2.5$ and $\tilde{\eta}_{12}=0.08$ (solid line); $\tilde{\eta}_1=1$, $\tilde{\eta}_2=1$ and $\tilde{\eta}_{12}=0$ (dashed line); $\tilde{\eta}_1=5.1$, $\tilde{\eta}_2=1.3$ and $\tilde{\eta}_{12}=0.7$ (dotted line). The plots (d)-(f) correspond to the fixed viscosity $\tilde{\eta}_1=0.6$, $\tilde{\eta}_2=2.5$ and $\tilde{\eta}_{12}=0.08$ and varying anisotropy: $\lambda_{\varepsilon}=(1.4)^{-1}$, $\lambda_{\sigma}=1$ (solid line); $\lambda_{\varepsilon}=1$, $\lambda_{\sigma}=1.4$ (dashed line).}\label{Velocity_plot}
\end{center}
\end{figure}

In this subsection we consider two similar types of the director arrangement $\mathbf{n}=(\cos\theta(y),\,\sin\theta(y))$ with
\begin{gather}\label{Patterns_BC}
\theta^B = \frac{\pi}{2} - \arcsin(\sin\pi y),\\
\theta^C =  \arcsin(\sin\pi y),
\end{gather} 
which we will refer to as the pattern (or cell) $B$ and $C$, respectively.
Similar to the pattern considered above, these patterns are also periodic, but with the period that is two times larger and consists of two different stripes.
Note that Eq.~\eqref{Patterns_BC} implies that the function $\arcsin (x)$ is restricted to its principal branch $\left[-\frac{\pi}{2}, \frac{\pi}{2}\right]$.
Because of this constraint, the direct calculation of \eqref{Concentration} results in a function $r_1$ that appears to have jump discontinuities at $y=\frac{1}{2}+k$, $k\in\mathbb{Z}$, contradicting the requirement that concentrations of ions have to be continuous.

This discrepancy is easily resolved once we observe that large gradients of ions concentrations are possible in our system due to the fact that the parameter $F\gg1$. Indeed, $F$ is equal to the ratio of electrostatic energy of an ion to its thermal energy so that the electrostatic forces dominate over diffusion when this parameter is large. It follows then that ions can pile-up in certain regions of the nematic electrolyte under the action of the field. Mathematically, this fact manifests itself in the presence of the boundary layers where this pile-up takes place and the discontinuous branches of $r_1$ that are obtained from the outer asymptotic solution \eqref{Concentration} need to be connected via an inner solution of \eqref{Eq_for_r_app} inside each boundary layer.

Briefly, if $y_0\in \frac{1}{2}+\mathbb{Z}$ is one of the points of discontinuity, let $y=y_0+\sqrt{\delta}\zeta$ to be the inner boundary layer variable and set $R(\zeta)=r(\sqrt{\delta}\zeta)$. Using the same expansions as in \eqref{Linearization} and collecting the leading order terms, we find that $R_1$ in the expansion of $R$ satisfies the following problem
\begin{equation}
\label{eq:inner}
\left\{
\begin{array}{ll}
 \left(1+\Delta_{\varepsilon}n_y(y_0)^2 \right)R_1^{\prime\prime}-2b\,R_1=2b\,c_1-M^\prime\left(y_0^+\right), &  \zeta>0,   \\
 \left(1+\Delta_{\varepsilon}n_y(y_0)^2 \right)R_1^{\prime\prime}-2b\,R_1=2b\,c_1-M^\prime\left(y_0^-\right), &  \zeta<0,   \\
 R_1\left(0^-\right)=R_1\left(0^+\right)\mbox{ and } R_1^\prime\left(0^-\right)=R_1^\prime\left(0^+\right), &       
\end{array}
\right.
\end{equation}
where $f\left(x^\pm\right)$ denote the right- and left-hand limits of $f$ at $x$, respectively. Further, the second equation in \eqref{Linearization} remains unchanged because the contribution of the boundary layer to the integral in \eqref{eq:ingr} appears at order $O(\delta^2)$. Hence $c_1=0$ and the solution to \eqref{eq:inner} is
\begin{equation}
\label{eq:inner_sol}
R_1(\zeta)=\left\{
\begin{array}{ll}
 \frac{1}{2b}M^\prime\left(y_0^+\right)+\frac{1}{4b}\left(M^\prime\left(y_0^-\right)-M^\prime\left(y_0^+\right)\right)e^{-\frac{\sqrt{2b}\zeta}{{\left(1+\Delta_{\varepsilon}n_y(y_0)^2 \right)}^{1/2}}}, &  \zeta>0,   \\
 \frac{1}{2b}M^\prime\left(y_0^-\right)+\frac{1}{4b}\left(M^\prime\left(y_0^+\right)-M^\prime\left(y_0^-\right)\right)e^{\frac{\sqrt{2b}\zeta}{{\left(1+\Delta_{\varepsilon}n_y(y_0)^2 \right)}^{1/2}}}, &  \zeta<0.   \\
\end{array}
\right.
\end{equation}
Note that this solution matches the branches of the outer solution to the right and to the left of $y_0$ in \eqref{Concentration} as $\zeta\to\pm\infty$ and it reduces to a constant value corresponding to \eqref{Concentration} evaluated at $y_0$ if $M^\prime$ is continuous at $y_0$. For simplicity, in what follows we will not present the expressions for the boundary layers solutions corresponding to particular patterns, although these solutions will be used in plotting of various fields.

Omitting the details, we use \eqref{Concentration} to write down the electric charge distribution outside of the boundary layers
\begin{equation}
Q^B = \begin{cases}
-\frac{\pi(\lambda_{\varepsilon}-\lambda_{\sigma})}{2G}\,\frac{1-\lambda_{\sigma}-(1+\lambda_{\sigma})\cos 2\pi y}{\left( \lambda_{\sigma}-(\lambda_{\sigma}-1)\sin^2\pi y\right)^2},\qquad |y|\,\text{mod}\,2\in\left[0, \frac{1}{2}\right]\cup [\frac{3}{2}, 2),\\
\linebreak\\
\frac{\pi(\lambda_{\varepsilon}-\lambda_{\sigma})}{2G}\,\frac{1-\lambda_{\sigma}-(1+\lambda_{\sigma})\cos 2\pi y}{\left( \lambda_{\sigma}-(\lambda_{\sigma}-1)\sin^2\pi y\right)^2},\qquad |y|\,\text{mod}\,2\in\left[\frac{1}{2}, \frac{3}{2}\right]
\end{cases}
\end{equation}
\begin{equation}
Q^C  =
\begin{cases}
-\frac{\pi(\lambda_{\varepsilon}-\lambda_{\sigma})}{2G}\,\frac{\lambda_{\sigma}-1-(1+\lambda_{\sigma})\cos 2\pi y}{\left( 1+(\lambda_{\sigma}-1)\sin^2\pi y\right)^2},\qquad |y|\,\text{mod}\,2\in\left[0, \frac{1}{2}\right]\cup [\frac{3}{2}, 2),\\
\linebreak\\
\frac{\pi(\lambda_{\varepsilon}-\lambda_{\sigma})}{2G}\,\frac{\lambda_{\sigma}-1-(1+\lambda_{\sigma})\cos 2\pi y}{\left( 1+(\lambda_{\sigma}-1)\sin^2\pi y\right)^2},\qquad |y|\,\text{mod}\,2\in\left[\frac{1}{2}, \frac{3}{2}\right]
\end{cases}
\end{equation}
for the pattern $B$ and $C$, respectively.

Similarly to $Q^A$, the charge concentrations $Q^B$ and $Q^C$ are proportional to the difference between $\lambda_{\varepsilon}$ and $\lambda_{\sigma}$ so that larger magnitudes of $Q^B$ and $Q^C$ should be observed in more anisotropic electrolytes. But unlike the pattern considered in the previous subsection, for the patterns $B$ and $C$, each period consists of two halves in which both positive and negative charges experience equivalent but alternating electrostatic forces. The resulting electro-osmotic flows then also exhibit an alternating pattern. This behavior occurs for any combination of the nematic viscosities, regardless of the degree of anisotropy (see Fig.~\ref{Velocity_plot}).

\begin{figure}
\begin{center}
\includegraphics[width=.4\textwidth]{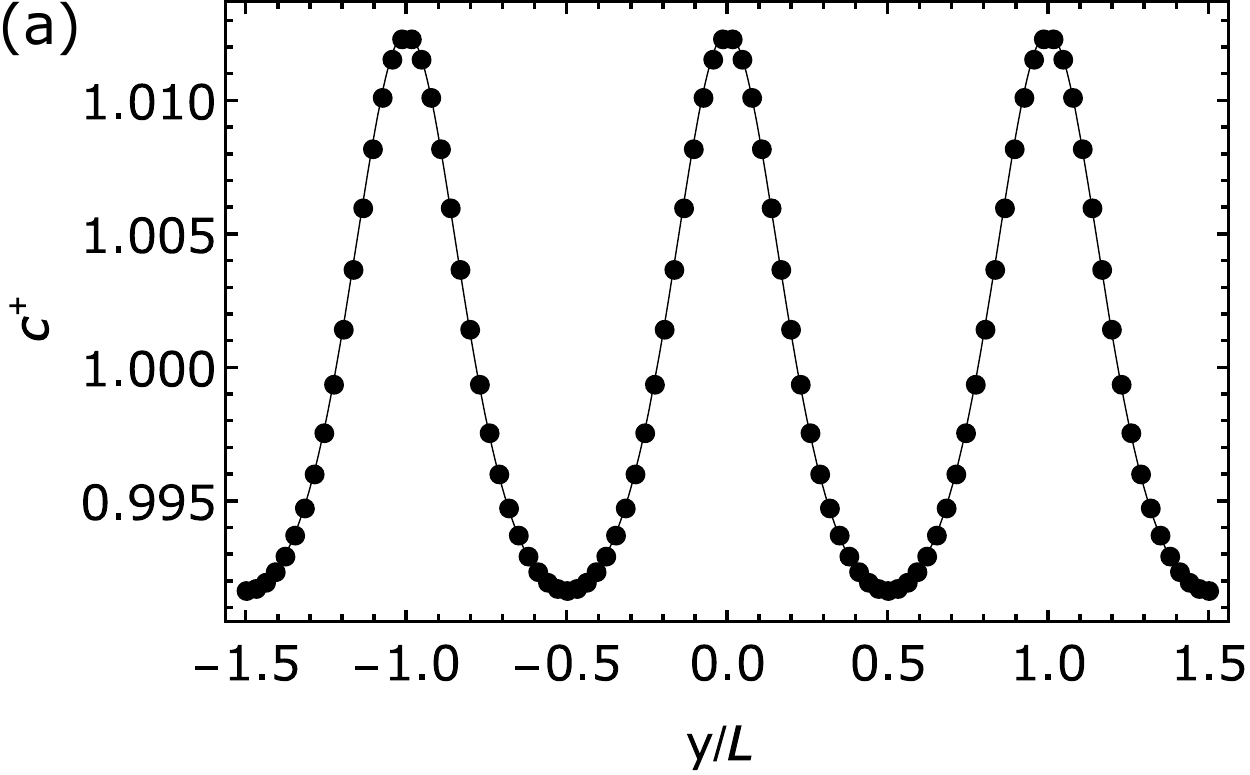}
\qquad \includegraphics[width=.4\textwidth]{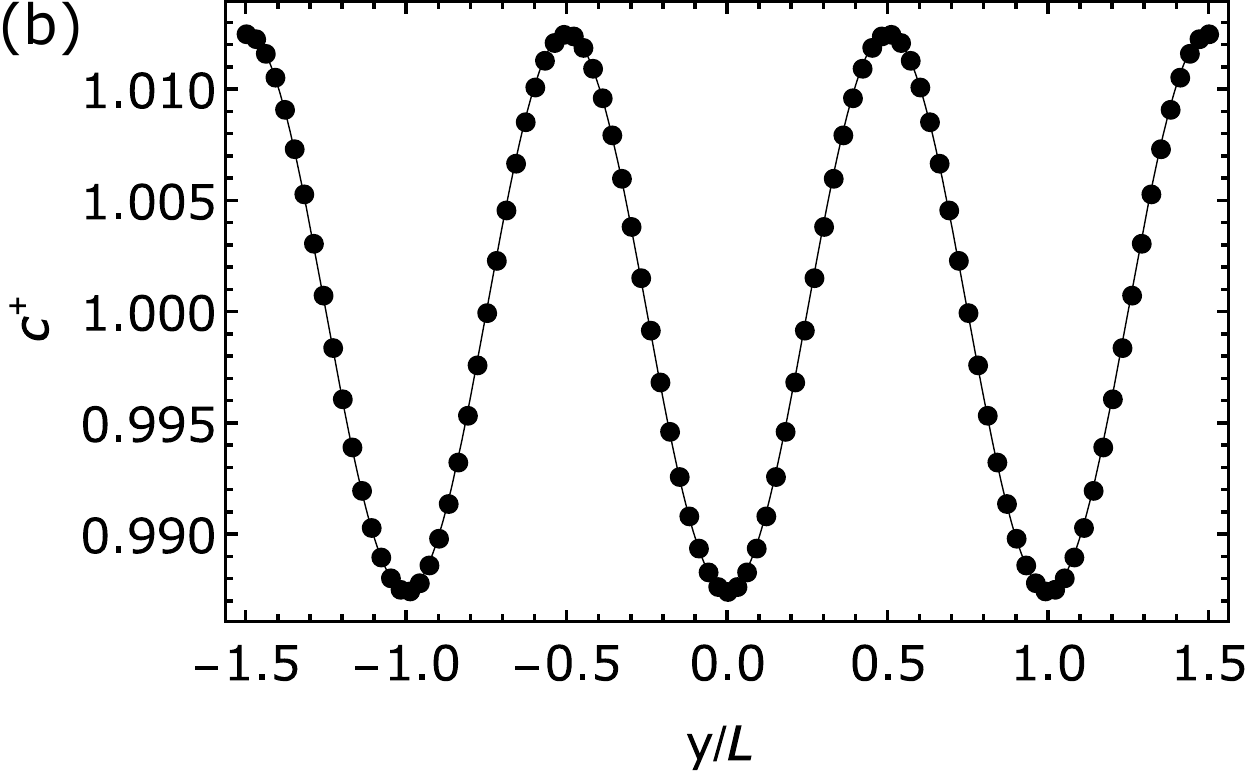}
\caption{Nondimensional concentration of the positive ions in the cell $A$ found from \eqref{Eq_for_r}-\eqref{Cond_for_r} numerically (points) and analytically (solid lines).
(a): $\lambda_{\varepsilon}=1$, $\lambda_{\sigma}=1.5$.
(b): $\lambda_{\varepsilon}=1.5$, $\lambda_{\sigma}=1$. 
Here $F=47$ and $G=63$ correspond to the experiment \citep{Peng_pattern}, where $E = 24$~mV$/\mu$m, $\bar{c} = 10^{19}$~m$^{-3}$, $L=50$~$\mu$m and $\varepsilon_{\perp}=6$.}\label{Numer_good}
\end{center}
\end{figure}

\subsection{Applicability of asymptotic solutions}

\begin{figure}
\begin{center}
\includegraphics[width=.4\textwidth]{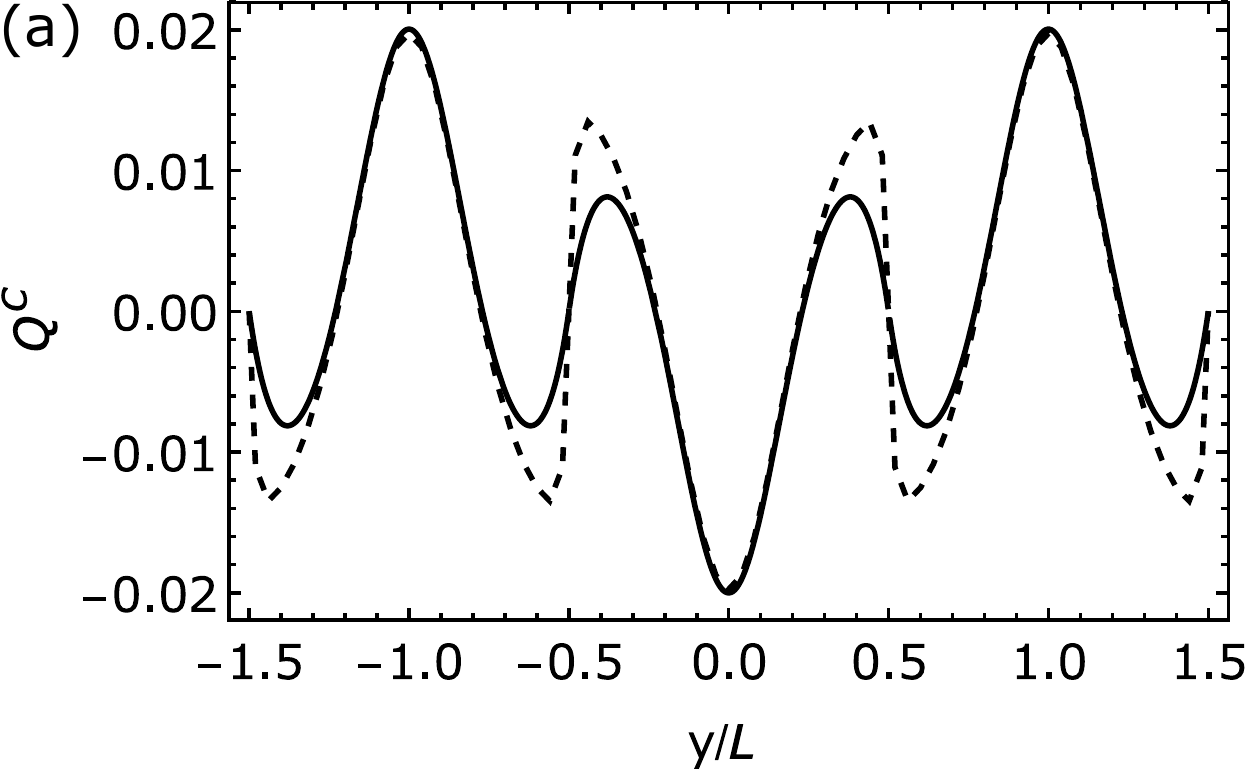}
\qquad \includegraphics[width=.4\textwidth]{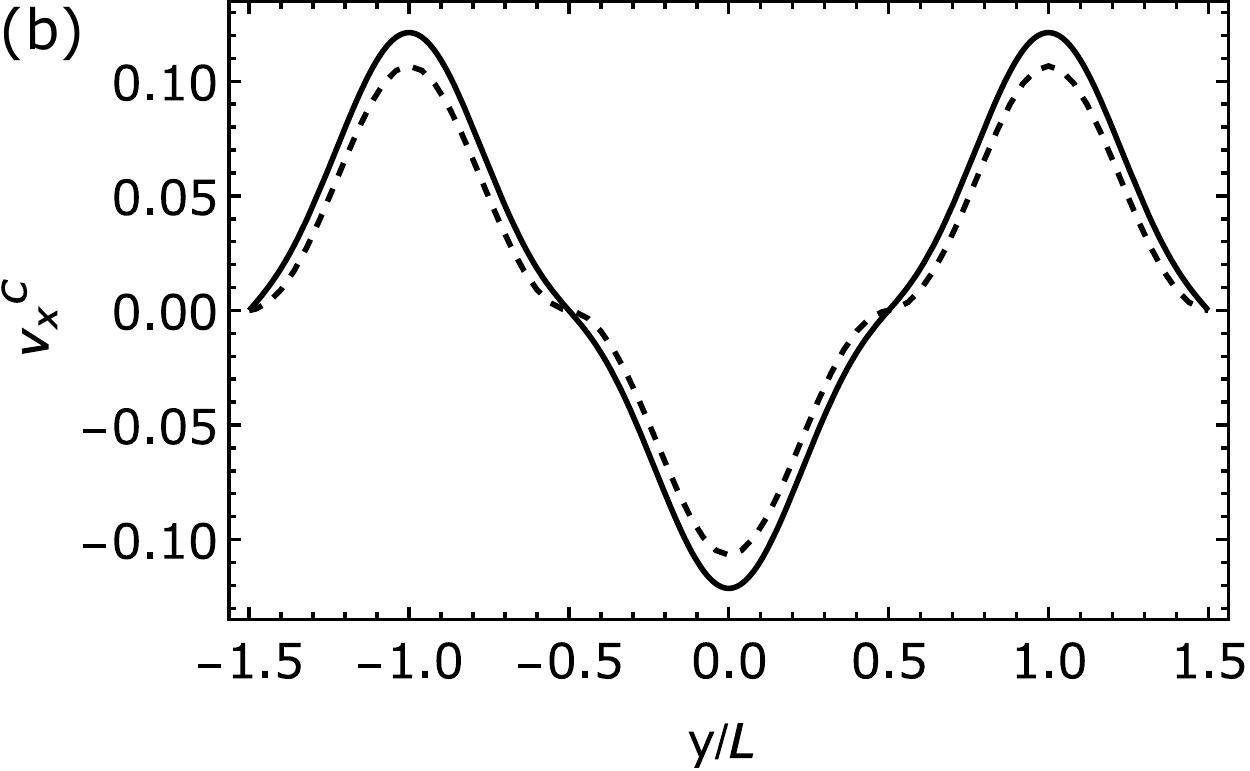}
\caption{Comparison between the numerical (dashed line) and asymptotic (solid line) solutions in the cell $C$. Here (a) depicts nondimensional charge concentration $Q^C=c^+-c^-$ and (b) depicts the nondimensional flow velocity. Both plots were obtained under the assumption that $\eta=1$, $\lambda_{\varepsilon}=1$, $\lambda_{\sigma}=1.4$ and $F=47$ and $G=63$.}\label{Numer_diffr_C}
\end{center}
\end{figure}

In the case of a prescribed quasi one-dimensional director field and equally charged and mobile cations and anions, the system of eleven governing equations \eqref{The_System} reduces to a single ordinary differential equation \eqref{Eq_for_r} for the function $r = \ln c^{+}$.
Supplemented by the integral constraint \eqref{Cond_for_r}, this equation allows us to find the spatial dependence of the ionic concentration and, subsequently, calculate all of the unknown quantities without the need to make any additional assumptions. Although the nonlinear Eq.~\eqref{Eq_for_r} is not solvable analytically, it can be solved asymptotically in the parameter regime corresponding to a typical nematic electrolyte. As shown in Fig.~\ref{Numer_good}, for the pattern $A$ the asymptotic solution \eqref{Concentration} is in excellent agreement with the results of numerical integration of the exact problem \eqref{Eq_for_r}-\eqref{Cond_for_r}. The leading order asymptotic approximation for the pattern $C$ shown in Fig.~\ref{Numer_diffr_C} is less accurate for the charge concentration within the boundary layer. However, even in this case the flow velocity agrees well with the numerical solution. Overall, the error incurred by using the approximate expression \eqref{Concentration} is quite sensitive to the values of the nondimensional parameters $\lambda_{\varepsilon}$, $\lambda_{\sigma}$, $G=\bar{c}eL/(\varepsilon_0\varepsilon_{\perp}E)$ and $F=eEL/(k_B\Theta)$.

A detailed discussion of the role of $G$ and $F$ can be found in \cite{Carme}. Briefly, the solution \eqref{Concentration} is correct as long as the following three conditions are satisfied: (a) The applied field $E$ is strong enough to overcome thermal fluctuations on the length scale of the stripe's width, i.e., $F\gg 1$ and diffusion can be neglected; (b) The applied field $E$ is lower than $E_c =\bar{c}eL/(\varepsilon_0\varepsilon_{\perp})$, i.e., $G\gg 1$; and (c) The anisotropy parameters $\lambda_{\varepsilon}$ and $\lambda_{\sigma}$ are finite and bounded away from zero. Here the critical value $E_c$ corresponds to a field that fully separates all the charges in the system so that the flow is no longer quadratic in the field strength.

\begin{figure}
\begin{center}
\includegraphics[width=.4\textwidth]{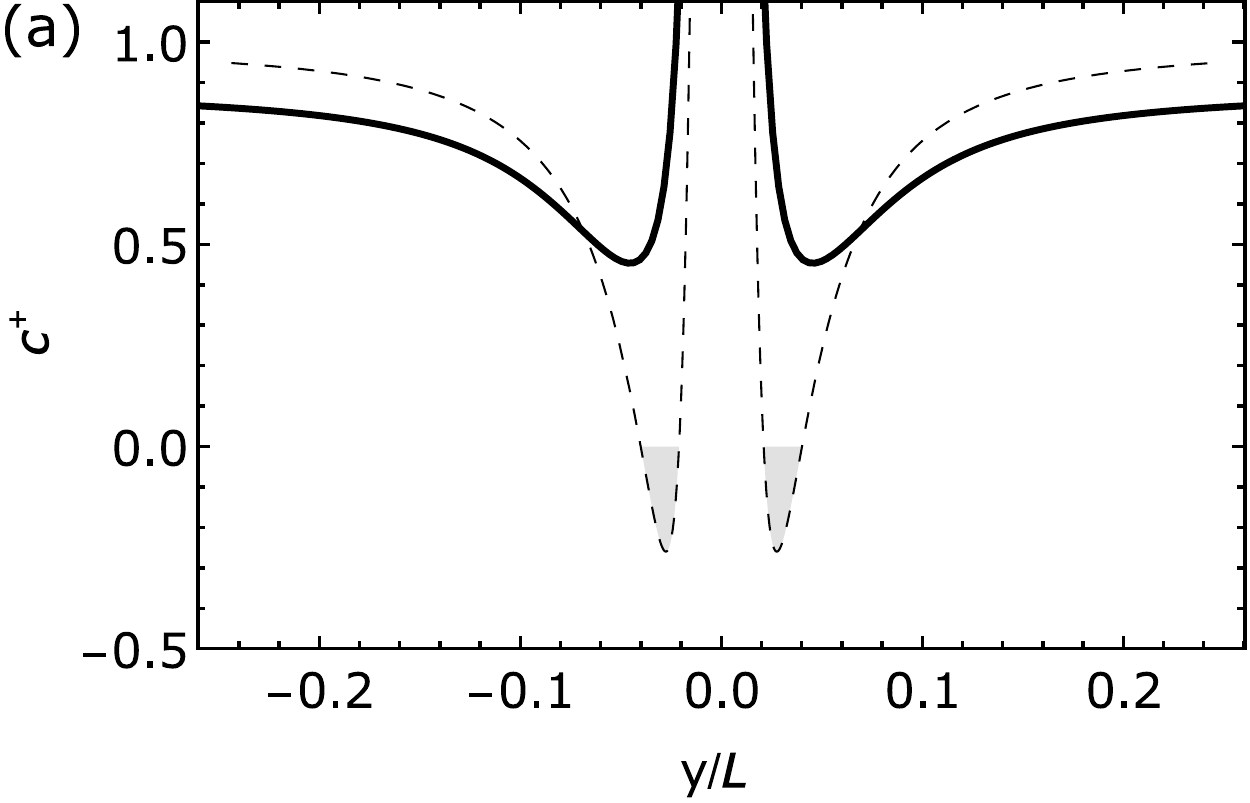}
\qquad \includegraphics[width=.4\textwidth]{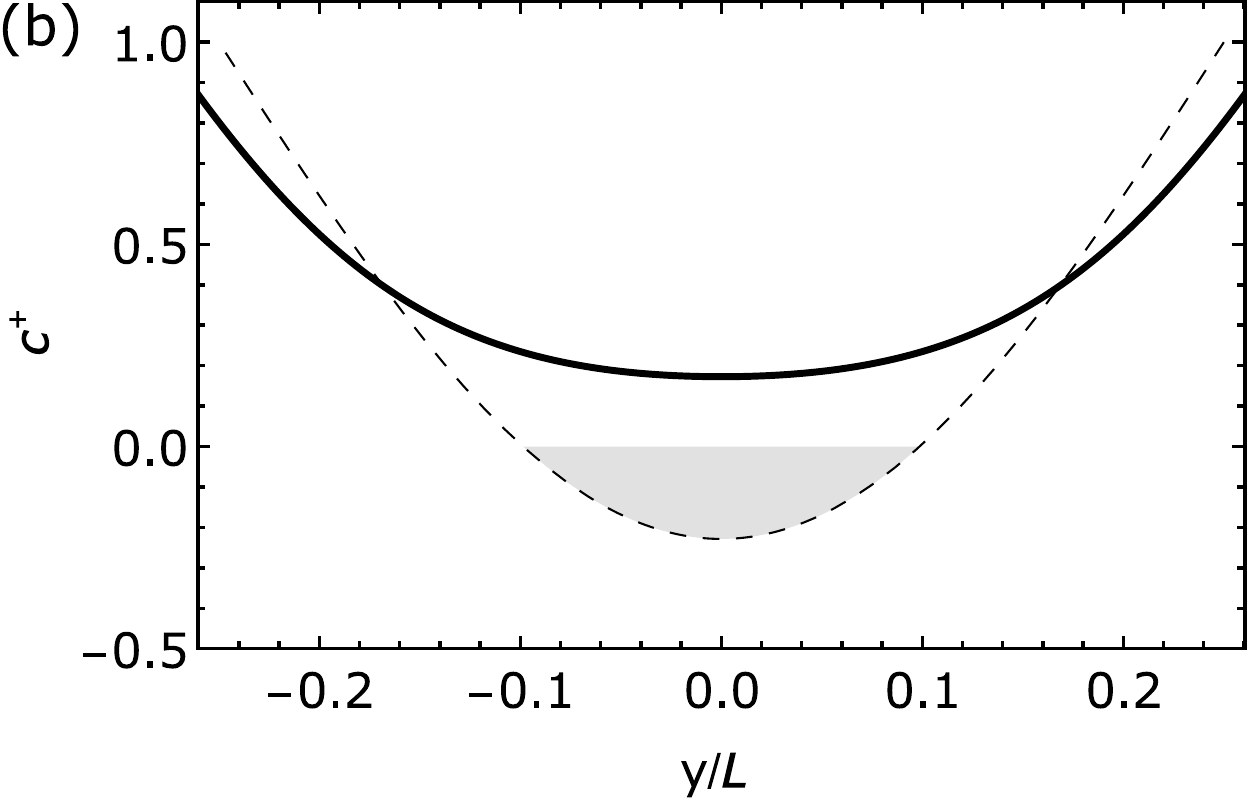}
\caption{Behavior of the nondimensional concentration $c^+$ found numerically (solid line) and analytically (dashed line) in the cell $A$ when the nematic is extremely anisotropic. Here (a): $\lambda_{\varepsilon}=1$, $\lambda_{\sigma}=400$; and (b): $\lambda_{\varepsilon}=50$, $\lambda_{\sigma}=1$. Both plots were obtained under the assumption that $F=47$ and $G=63$.}\label{Numer_diffr}
\end{center}
\end{figure}

To illustrate the role of the assumption (c) that was not explicitly discussed in \cite{Carme}, it is convenient to consider the pattern $A$. Indeed, in this case the corresponding expression for $c^+$ is defined in a simpler way as can be seen from \eqref{Concentartion_A} while recalling that $c^{\pm}=1\pm\frac{1}{2}Q$.
One can easily find that if $\lambda_{\sigma}\to\infty$, then $c^+ \to 1-\pi\csc^2\pi y/(2G)$.
Similarly, $c^+ \to 1-\pi\lambda_{\varepsilon}\sec^2\pi y/(2G)$ when $\lambda_{\sigma}\to 0$.
Clearly, both of these expressions result in negative concentrations for certain values of $y$ (see Fig.~\ref{Numer_diffr}).
The reason for this can be traced back to the equation \eqref{Linearization} that was derived under the assumption that the deviations of $c^+$ from the bulk concentration $\bar{c}$ are small and $c^{+}=e^r\approx 1+\delta r_1$.
But this is not the case when $\lambda_{\sigma}\gg 1$ or $\lambda_{\sigma}\ll 1$ because then $r_1$ is not small everywhere.   
Note that, even though the asymptotic analysis fails in these regimes, a numerical solution of the problem \eqref{Eq_for_r}-\eqref{Cond_for_r} predicts concentrations correctly (cf. Fig.~\ref{Numer_diffr}).

In a similar manner, one can show that the solutions \eqref{Concentration} are incorrect in the case of strong positive dielectric anisotropy, $\lambda_{\varepsilon}\gg 1$, since $c^{\pm}$ grows linearly with $\lambda_{\varepsilon}$.
Mathematically this is because the first term in Eq.~\eqref{Eq_for_r} is no longer small and cannot be neglected.

It should be noted that the asymptotic solutions \eqref{Concentration} give qualitatively improper results only in extremely anisotropic nematics, in which mobilities or dielectric permittivities differ by orders of magnitude.
Otherwise the approximate expressions \eqref{Concentration} are quite accurate.
Even in the case of $\lambda_{\varepsilon}=10$ or $\lambda_{\sigma}=10$ they deviate from the numerical results by $\approx 5\%$.  

\section{Conclusions}

We considered a nematic electrolyte, an ideal ionic gas in the liquid crystalline matrix, and proposed a theoretical model of electro-osmosis in such a medium.
We showed how the equations governing this phenomenon can be derived in a simple and efficient way from a variational principle of the least energy dissipation.
An advantage of the proposed approach  is that it can be easily reformulated in terms of the tensorial order parameter instead of the director.
This feature opens a way to a theoretical description of electrophoretic transport of colloidal inclusions, which are typically accompanied by topological defects.

As an illustrative example, electro-osmotic flows in nematic films with prescribed periodic molecular orientation were considered.    
Even this quasi one-dimensional problem cannot be exactly solved analytically.
Its asymptotic solutions, however, are in good agreement with the results of experiments and numerical simulations.

The proposed approach clearly demonstrates that the necessary condition for liquid-crystal-enabled electro-osmosis is a spatially varying electric charge density $Q(\mathbf{r})\propto E$.
The directed motion of the charges under the action of the electric force $QE\propto E^2$ results in the flow of the liquid crystal.
Since the driving force is quadratic in $E$ the velocity of the flow does not depend on the field's polarity.

The charge density $Q(\mathbf{r})$ itself arises from an interplay between non-uniform director field and anisotropic properties of its dielectric permittivity $\hat{\mathbf{\varepsilon}}$ and conductivity $\hat{\mathbf{\sigma}}$.
The latter two can mutually suppress as well as enhance each other.
Hypothetically, in nematics with varying $\hat{\mathbf{\sigma}}$ and $\hat{\mathbf{\varepsilon}}$ this competition can be exploited for dynamic switching of the flow's direction.

\begin{acknowledgments}
The authors acknowledge support from NSF DMS-1434185.
NJW was also supported in part by National Science Foundation Grant DMS-1418991.
\end{acknowledgments}

\bibliography{e-osmosis}

\end{document}